\documentclass[12pt]{article}

\usepackage[utf8]{inputenc} 
\usepackage[T1]{fontenc}    
\usepackage{hyperref}       
\usepackage{url}            
\usepackage{booktabs}       
\usepackage{amsfonts}       
\usepackage{nicefrac}       
\usepackage{graphicx}
\usepackage{authblk}
\usepackage{titlesec}
\usepackage{natbib}
\usepackage{amsmath}

\setcitestyle{authoryear,round,semicolon}

\usepackage{caption, setspace}
\DeclareCaptionFont{helv}{\normalsize\fontfamily{phv}\selectfont}

\usepackage{geometry}
\title{\normalsize
Do biological constraints impair dendritic computation?
}

\author[1,*]{Ilenna Simone Jones}
\author[1,2]{Konrad Paul Kording}
\affil[1]{Department of Neuroscience, University of Pennsylvania}
\affil[2]{Department Bioengineering, University of Pennsylvania}
\affil[*]{Corresponding author: Ilenna Jones, ilennaj@pennmedicine.upenn.edu}

\begin{document}

\maketitle
\date{}
\begin{quote} 	{\normalfont\normalsize\bfseries\centering Abtract\par}
\indent
Computations on the dendritic trees of neurons have important constraints. Voltage dependent conductances in dendrites are not similar to arbitrary direct-current generation, they are the basis for dendritic nonlinearities and they do not allow converting positive currents into negative currents. While it has been speculated that the dendritic tree of a neuron can be seen as a multi-layer neural network and it has been shown that such an architecture could be computationally strong, we do not know if that computational strength is preserved under these biological constraints. Here we simulate models of dendritic computation with and without these constraints. We find that dendritic model performance on interesting machine learning tasks is not hurt by these constraints but may benefit from them. Our results suggest that single real dendritic trees may be able to learn a surprisingly broad range of tasks.  
\end{quote}

\newpage

\section*{Introduction}

The dominating point-neuron idea \citep{McCulloch1943} which underlies artificial neural networks (ANNs) \citep{Lecun2015, Krizhevsky2012, Minsky1969} assumes that dendrites are linear integrators of synaptic inputs. However dendrites are highly nonlinear due to their voltage dependent ion channels \citep{Antic2010}. Detailed models of dendritic processing thus use multi-compartment modeling with differing levels of abstraction \citep{Hines1997,Izhikevich2007} allowing for good models for phenomena such as direction sensitivity \citep{Rall1964,Koch1982} and coincidence detection \citep{Agmon-Snir1998, London2005}. However, we do not just want to ask if a realistic neuron can do one thing, but also to understand the general class of problems a neuron can solve. In this vein, the perceptron learning rule has been used to train a biophysical  multi-compartment model to do binary classification tasks \citep{Moldwin2019}. We know that real neurons can solve interesting problems, e.g. the XOR task \citep{Gidon2020}, which requires at least two layers in an ANN. Indeed, recent computational work \citep{David2019} has shown that  a single dendritic tree can only be simulated with a many layered neural network of limited width. From these, it is clear that a dendrite can do more than linearly integrate, but we must also consider that there are important biological constraints. This raises the question of how the biological properties of dendrites affect their computational capabilities. 

Owing to their rich nonlinear properties, real dendrites have been seen as equivalent to a multi-layer ANN \citep{Poirazi2003,Poirazi2003a,Mel2016,Poirazi2020}. Specifically, it is generally seen as a multi-layer ANN that is constrained to have a tree structure where inputs are successively combined with one another on the way to the soma where the spike generator sits. Inspired by these ideas, our own past work modeled a single neuron as using a binary tree structure \citep{Jones2021}. These ANN neuron models utilize one important biological constraint, the tree structure of the dendritic tree, and test its impact on neuronal output and computation respectively. However, these studies only looked at a small set of biological constraints.

This lack of biological constraints results in models abstracted away from biological reality. Here we focus on three of these abstractions in our own tree-structured single neuron model \citep{Jones2021} and identify ways it could be more biologically plausible.
First, its nonlinear activation function is a Leaky Rectified Linear Unit (LReLU). Though LReLU may be a widely used activation function in many successful deep ANNs \citep{Nair2010,Krizhevsky2012,Xu2015, Maas2013}, perhaps we can rely on voltage-dependent ion channel nonlinearities to identify a better activation function. Second, its inputs are directly received by the model as values between 0 and 1. Perhaps we can make input more realistic by mapping this input through a conductance-based synaptic nonlinearity in the units of millivolts \citep{Koch1999}. Third, its weight parameters can be both positive and negative, allowing the activations traveling through the tree to flip signs. The weights in the binary tree structure of our model would be analogous to axial conductance between compartments \citep{Hines1997}, and so an improvement could be to constrain these weights to be non-negative \citep{Rall1967}. These are ways we can move a highly abstracted model closer toward biological realism. However, we suspect that imposing further biological constraints may yield differing performance results.

Here we use these suggested improvements as three biological constraints for our binary tree model: synaptic input nonlinearities, voltage-gated-ion-channel-derived dendritic nonlinearities, and non-negative axial conductance weights. We observe the impacts of introducing each constraint by training and testing the binary tree model on 7 different binary classification machine learning datasets. We find that the dendritic nonlinearity activation function introduced performs as well if not better than other typical activation functions. We find that synaptic input constraints generally do not limit or expand computation performance. Interestingly we see that the non-negative weight constraint limits performance in low multi-synaptic-input cases, but the combination of this non-negative weight constraint and the synaptic input constraint eliminates this limitation. Lastly, we find that the binary tree model with all three constraints performs as well or better than the model without these constraints. In conclusion, implementing these constraints fulfills the dual goal of creating a more biologically realistic binary tree neuron model and showing us that these combined constraints don't limit, and may actually improve, the model's computational capability. This suggests real dendritic trees may be able to perform a broad range of complex functions.

\subsection*{Background}

\subsubsection*{Local dendritic dynamics}
Neuroscience has a rich set of methods for the simulation of neurons, based on cable equations with an addition of a host of neuron-specific knowledge. The resulting simulations can generate the time-varying activities of a neuron in response to arbitrary stimuli. Here we use an approximation that is different in two ways. First, we only want to consider constant inputs. Second, we only want to ask if the voltage will keep going up and a spike will result and we are not interested in the dynamics of the spike itself. For those cases we will make a steady state approximation for currents going from the periphery towards the soma. This dendritic nonlinearity or activation function, which we call the NaCaK function (as it is meant to capture sodium, calcium, and potassium currents), is but one of many potential approaches to representing the nonlinear local dendritic dynamics.

When we have a node in the neuron there will be currents coming in from the parent dendritic branches (which have currents $I_1$ and $I_2$), we will have local sodium ($I_Na$), calcium ($I_Ca$), and potassium ($I_K$) currents, we will have the current relating to charging the local capacitance ($I_C$), and currents going towards the soma ($I_{out}$),

\begin{align}
    I_{total}=I_1+I_2 +I_{Na}+ I_{Ca}+ I_{K} + I_{C}- I_{out}.  \\ \nonumber \label{eq1}
\end{align}

For the maximal conductances ($\bar{g_{Na}}, \bar{g_{Ca}}, \bar{g_{K}}$) and current-voltage (IV) curves of sodium \citep{Hodgkin1952}, potassium \citep{Doiron2001}, and calcium \citep{Miyasho2001} channels there exists an extensive literature. We used IV curves that were derived from biophysical simulations of each ion channel using BRIAN2 \citep{Branco2010}, wherein the peak current was recorded for a range of voltage clamp settings. Given the current peaks for these ion channels occurred near instantaneously (within 1 millisecond) and we assume we are only dealing in local steady state dynamics. We then fit continuous functions to each of the IV curves, yielding the following equations,

\begin{align}
f_{Na}(V) & = \frac{0.0878 (113.68\text{mV} - V)}{1\text{mV} + \exp(6.39\text{mV} - V/8.98)} \\ 
f_{Ca}(V) & = \frac{0.129(69.62\text{mV} - V)}{1 + \exp(-4.40\text{mV} - V/4.25)} \\ 
f_{K}(V) & = \frac{-2.23\text{mV}}{0.436\text{mV} + \exp(-0.132(V-16.74\text{mV}))}.
\end{align}
We thus have a set of meaningful functions for local currents that form the NaCaK function.

The dynamics on the dendritic tree are complex and time-varying, and contain spiking dynamics. Instead of simulating such a complex model we will use a very simple approximation:

\begin{align}
    V \approx & \frac{1}{K} (g_1 V_1 + g_2 V_2 + \bar{g_{Na}} f_{Na}(V_0) + \bar{g_{Ca}} f_{Ca}(V_0) + \bar{g_{K}} f_{K}(V_0)). \\ \nonumber
\end{align}

We set $\bar{g_{Na}}, \bar{g_{Ca}}$, and $\bar{g_{K}}$ to equal 1, and $\frac{1}{K} = 0.5$ in our model.

Our dendritic nonlinearity function, the NaCaK function, takes on the role of the activation function in more traditional ways of conceptualizing dendrites as neural networks \citep{Poirazi2003a,Jones2021}. Our derivation of this result of course made problematic assumptions about how to solve the relevant implicit equations. Another way of seeing the NaCaK function is simply a way of conceptually capturing the nonlinear effects induced by the three channel types.

\subsubsection*{Local synaptic dynamics approximation derivation}
Just like in the case of the overall currents and local nonlinearities above, we will approximate the synapse properties, which are the input layer in our model. Using standard biophysical components, we can derive the steady state voltage by considering capacitative ($I_C$), membrane ($I_R$), excitatory ($I_e$), inhibitory ($I_i$), leak ($I_{leak}$), and "injected" ($I_{inj}$) currents,

\begin{align}
I_{total} = I_C + I_R + I_e + I_i + I_{leak} + I_{inj}. \\ \nonumber \label{eq10}
\end{align}

We can replace these current variables with their biophysical terms, while also ignoring membrane currents given the area of a synapse is small. For simplicity, both excitatory and inhibitory currents are included in the same synapse, so this formulation does not follow Dale's law. $I_{inj} = 0$ in this case as well,
\begin{align}
I_{total} = C \frac{dV}{dt} + g_e (V - E_e)  + g_i (V - E_i) + g_0 (V - E_0). \\ \nonumber \label{eq11}
\end{align}

The conductances $g_e$ and $g_i$ are functions of time and could be approximated from the slope of an I-V plot of peak currents of excitatory and inhibitory channels \citep{Koch1999}. The positive values of $g_e$ and $g_i$ will be calculated using a set of parameters $\alpha$ that characterize each synapse, can be trained and define the conductances:
\begin{align}
g_p(x) = \exp(\alpha_{1p}x + \alpha_{2p}). \\ \nonumber \label{eq12}
\end{align}

Where the input variable $x$, which is between 0 and 1, can be interpreted as probability of vesicular release. The parameter $\alpha_{1p}$ can be interpreted as an approximation of number of quantal release sites or vesicle size. The parameter $\alpha_{2p}$ can be interpreted as peak conductance as a result of triggering post-synaptic receptors, which would be a function of however many post-synaptic receptors are present. This formulation is similar to that which is elaborated on in \citet{Koch1999} (p. 91). However, it is different because it is an exponential expression, which we used to force the synaptic conductance to be positive and differentiable.  Lastly, $g_0$ is a trainable parameter, which means there are 5 trainable parameters for each synapse ($\alpha_{1e}, \alpha_{2e}, \alpha_{1i}, \alpha_{2i}$, and $g_0$). These components in this formulation add more synaptic parameters (degrees of freedom) replacing what is typically assumed to be a scalar value.

The steady state convergence voltage where $\frac{dV}{dt}=0$ can then be calculated:

\begin{align}
V_{\infty} = \frac{g_e E_e + g_i E_i + g_0 E_0}{g_e + g_i + g_0 }. \\ \nonumber \label{eq13}
\end{align}

We set $E_e$, $E_i$ and $E_0$ to be 50 mV, -70 mV and -65 mV to correspond to excitatory, inhibitory, and leak channels at the synapse.

We thus have a meaningful approximation of the effect of a synapse on the downstream dendrites.

\begin{figure}[ht]
\centering
\includegraphics[width=0.9\linewidth]{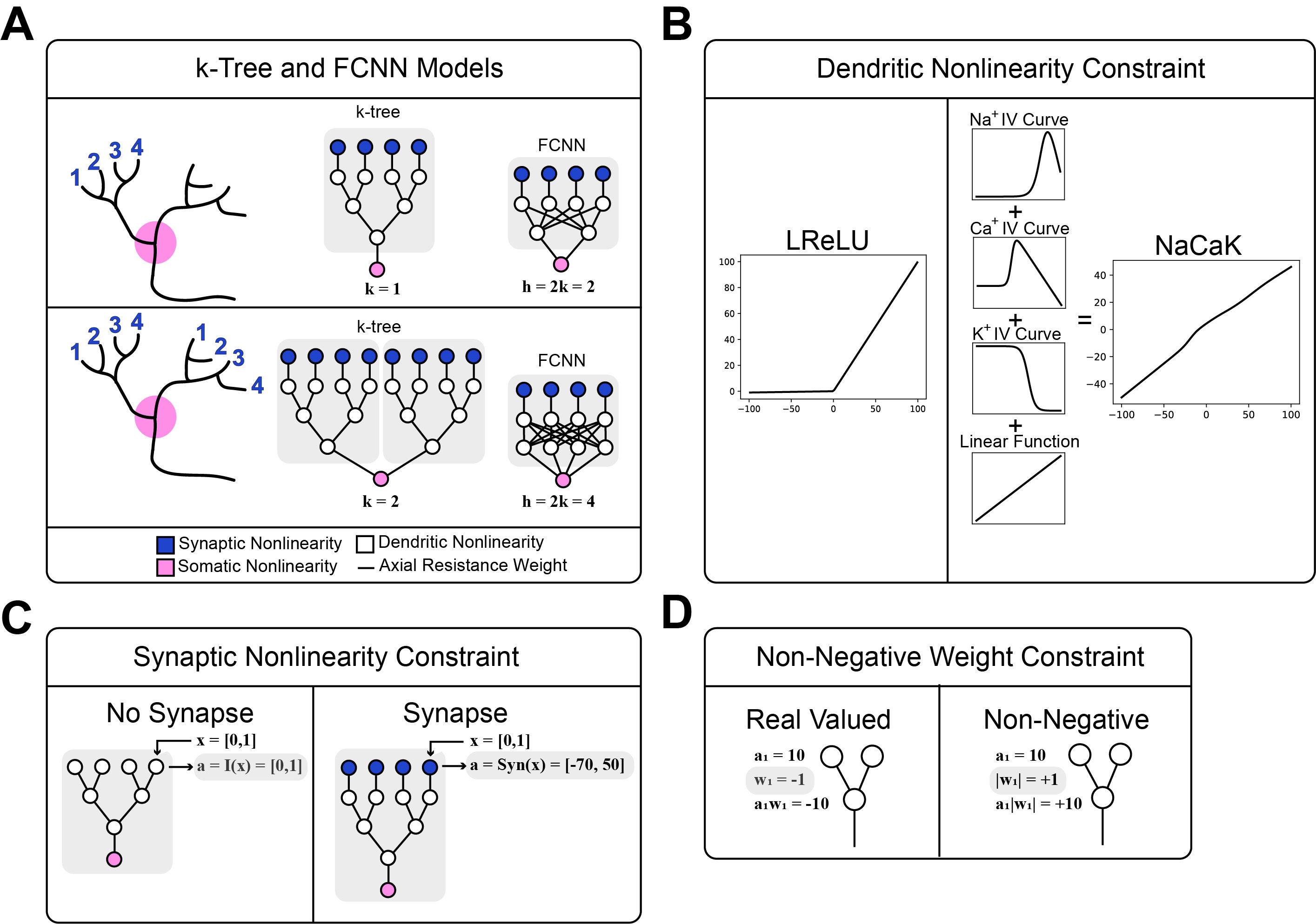}
\caption{Binary $k$-tree Model and Fully Connected Neural Network (FCNN) Control Model Containing Three Biological Constraints. A) Blue nodes apply an optional synaptic nonlinearity to the input. White nodes apply an optional dendritic nonlinearity. Black lines connecting the nodes are weights that are optionally constrained to be non-negative. The value $k$ determines how many sub-trees of identical input are repeated in the model as an analogy to repeated synaptic inputs on different dendritic trees. The value $h$ corresponds to the number of nodes in the hidden layer of the FCNN. B) Comparison of LReLU activation function and NaCaK activation function. C) Comparison of model without synapse nonlinearity and with synapse nonlinearity. D) Comparison of model weights without and with the non-negative weight constraint.}
\label{fig:model}
\end{figure}

\section*{Results}

In this paper we modify the previously established ANN dendritic neuron model, called the k-tree \citep{Jones2021}, and add three more dendritic constraints to it in a dual effort of making it more biologically plausible and observing the impact of these constraints on model performance when they are introduced. We introduce NaCaK nonlinearity (Figure \ref{fig:model}B, Figure \ref{fig:derivatives}),  which is a linear combination of sodium, calcium, and potassium IV curves that map voltage input to current output. We introduce a synapse nonlinearity (Figure \ref{fig:model}C), which is a steady state synaptic voltage response as a function of a continuous value that represents vesicular dynamics. This synapse nonlinearity output is in millivolts. Lastly, we introduce an axial conductance weight constraint (Figure \ref{fig:model}D) where the weight parameters between each node is constrained to be non-negative since in neuron models conductance is a scalar value. As a control, we use a parameter-size-comparable 2-layer Fully Connected Neural Network (FCNN) dense network to see the impact of the sparse binary tree structure. We also have multi-synaptic-input repetition conditions to see the impact of repeated inputs to different dendritic sub-trees. (Figure \ref{fig:model}A). Implementing these biological constraints and comparing their presence with their absence allows us to observe the computational impact of these constraints.

\begin{figure}[ht]
\centering
\includegraphics[width=0.9\linewidth]{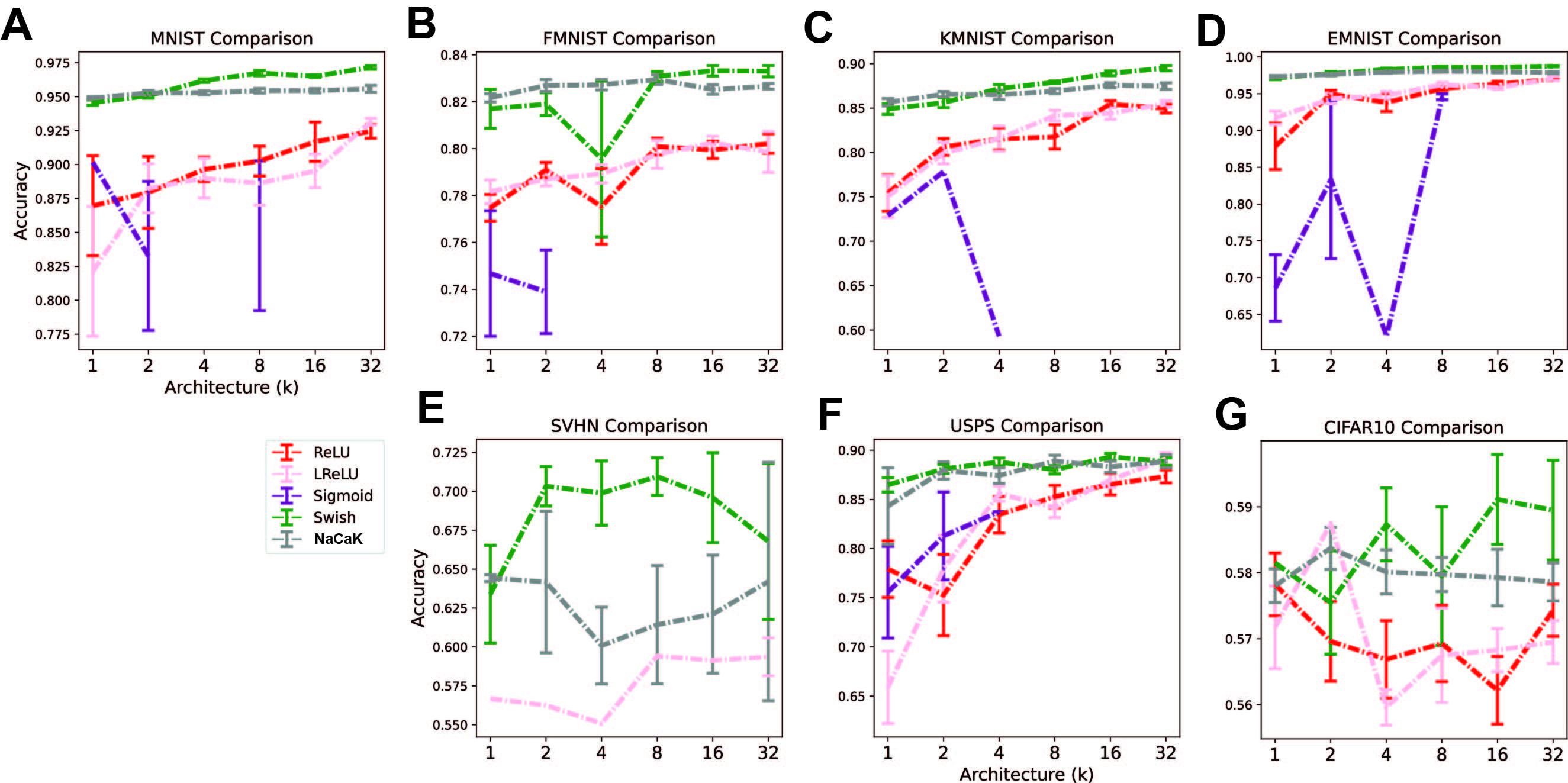}
\caption{Comparing Nonlinear Activation Functions for the Dendrite Nodes in the $k$-tree For Each Dataset. Architecture ranges from low input multi-synaptic input repetitions ($k$=1) to high input repetitions ($k$=32) and mean accuracy of each model is displayed with standard error bars. Number of trials ranges from 1-10, and trial accuracy was omitted if the training was deemed to fail (accuracy threshold 0.55). Table \ref{tab-ktree_nonlins} lists trial counts. Red: Rectified Linear Unit (ReLU), Pink: LReLU, Purple: Sigmoid, Green: Swish, Grey: NaCaK Nonlinearity}
\label{fig:k-treenonlins}
\end{figure}

\subsection*{NaCaK nonlinearity outperforms other nonlinearities in $k$-tree model}

In densely connected ANNs, Rectified Linear Unit (ReLU) and LReLU have been accepted to improve model performances. However since we are using a sparsely connected model architecture, the $k$-tree, we are interested in seeing if the NaCaK function differentially impacts $k$-tree model performance. To do this we tested the $k$-tree performance on various datasets with the NaCaK activation function and compare this performance to the ReLU, LReLU, Sigmoid, and Swish activation functions. We find that the NaCaK nonlinearity outperforms ReLU, LReLU, and sigmoid activation functions in 5 out of 7 datasets in the $k$-tree model (Figure \ref{fig:k-treenonlins}A-E, Figure \ref{fig:k-treenonlins_pval}A-E). Interestingly, the swish nonlinearity also performs about as well or better than the NaCaK nonlinearity in the $k$-tree model. (Figure \ref{fig:k-treenonlins}A-F, Figure \ref{fig:k-treenonlins_pval}A-F). For the FCNN model, ReLU and LReLU architectures performed better than all other nonlinearities in 4 out of 7 datasets (Figure \ref{fig:fcnn_nonlins}A-D,G, Figure \ref{fig:fcnn_nonlins_pval}A-D,G).  These results show that the NaCaK and swish functions paired with the binary tree architecture perform better than other nonlinearities. 

\begin{figure}[ht]
\centering
\includegraphics[width=0.9\linewidth]{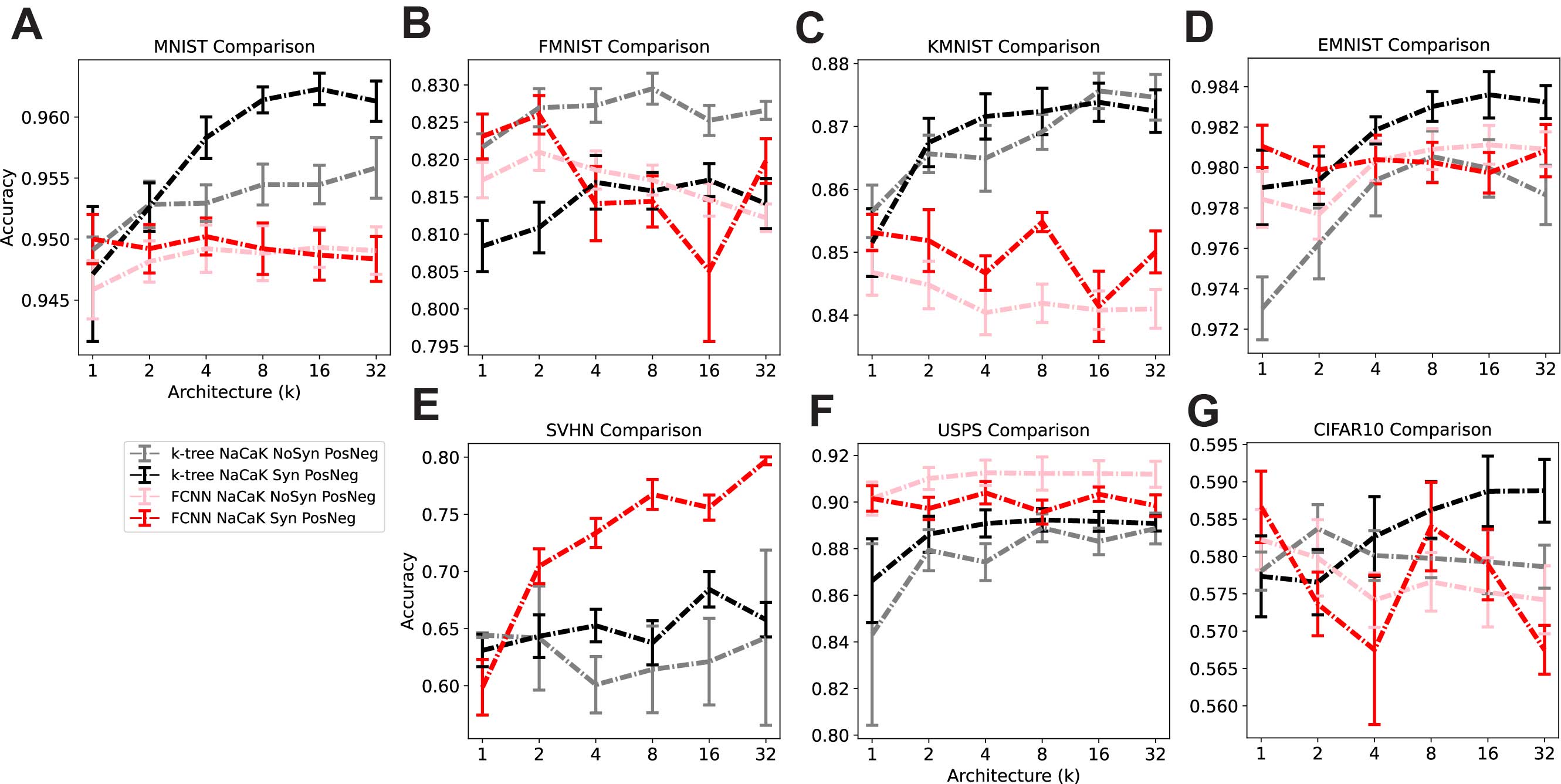}
\caption{Comparing $k$-tree and FCNN models With or Without Synapse nonlinearity. Architecture ranges from low input multi-synaptic input repetitions (k=1) to high input repetitions (k=32) and mean accuracy of each model is displayed with standard error bars. Number of trials ranges from 1-10, and trial accuracy was omitted if the training was deemed to fail (accuracy threshold 0.55). Table \ref{tab-synapse} lists trial counts. Red: FCNN with Synapses, Pink: FCNN without Synapses, Black: $k$-tree with Synapses, Grey: $k$-tree without Synapses.}
\label{fig:synapse}
\end{figure}

\subsection*{Addition of synaptic input nonlinearity can improve performance of binary tree model}

Without synapse nonlinearity mapping of the inputs, the $k$-tree model is receiving inputs that are not biologically realistic. We want to see if using the synapse nonlinearity constraint (Figure \ref{fig:model}C), which nonlinearly maps the input to realistic millivolt units, impacts $k$-tree model computation. In order to show this, we implement a steady-state voltage conception of synaptic dynamics in order to map the input values to synaptic voltage output. Every input corresponds to a differing synaptic nonlinearity function. We allow the synaptic nonlinearities to have 5 different trainable parameters ($\alpha_{1e}, \alpha_{2e}, \alpha_{1i}, \alpha_{2i}, g_0$) instead of simply 1 trainable weight. We thus compare the densely connected FCNN and the binary tree models under the synapse or no-synapse conditions, all using the NaCaK nonlinearity, in order to observe their impacts on computation.

\begin{figure}[ht]
\centering
\includegraphics[width=0.8\linewidth]{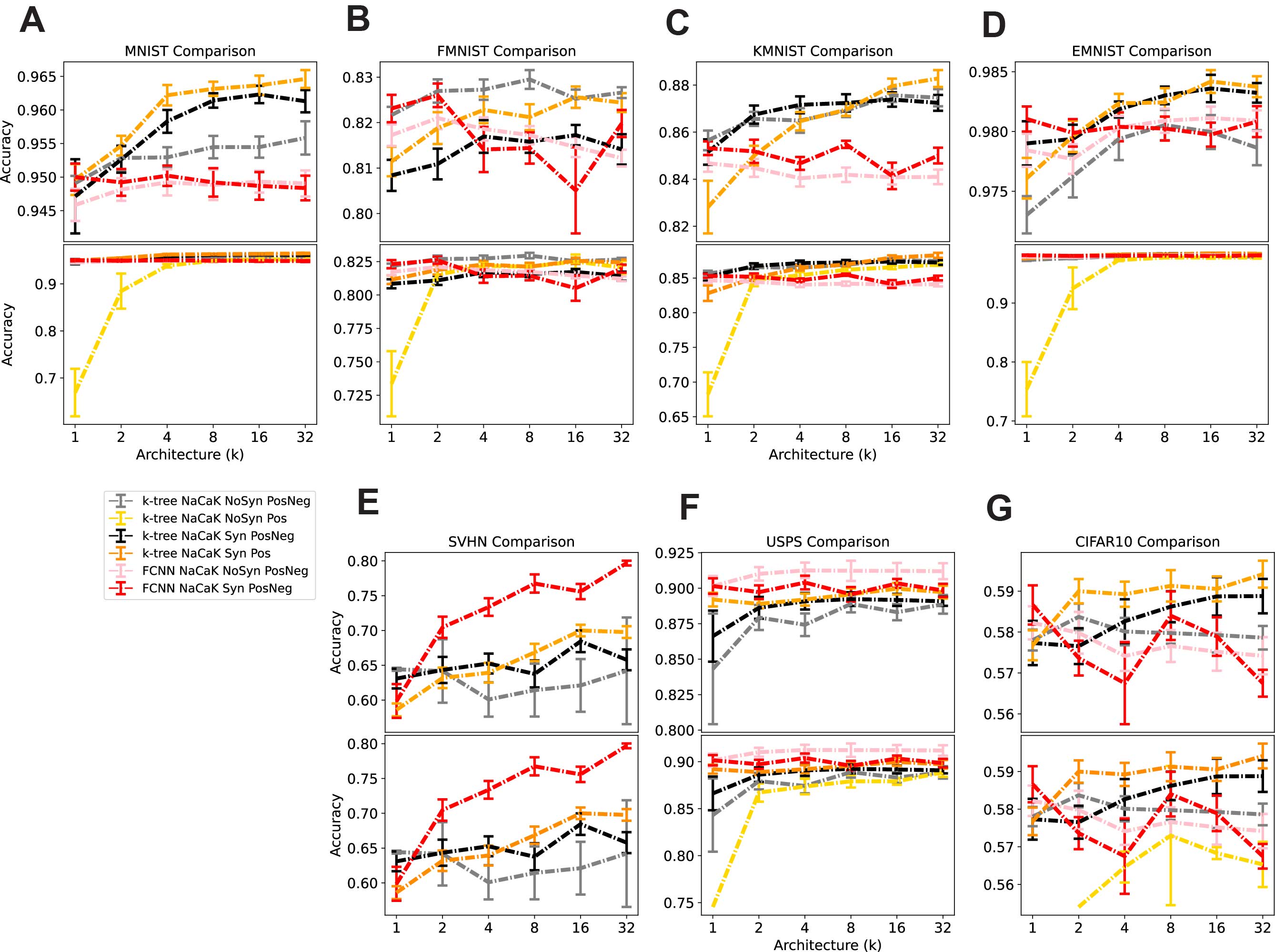}
\caption{Comparing $k$-tree and FCNN models With or Without the Non-negative Weight Constraint. Architecture ranges from low input multi-synaptic input repetitions (k=1) to high input repetitions (k=32) and mean accuracy of each model is displayed with standard error bars. Number of trials ranges from 1-10, and trial accuracy was omitted if the training was deemed to fail (accuracy threshold 0.55). Table \ref{tab-constantsqgl} lists trial counts. Orange: $k$-tree with Synapses and with Non-negative weights, Yellow: $k$-tree without Synapses and with Non-negative weights Red: FCNN with Synapses, Pink: FCNN without Synapses, Black: $k$-tree with Synapses, Grey: $k$-tree without Synapses. Top graphs in A-G omit the accuracies of the $k$-tree without synapses and with non-negative weights in order to resolve the lines with higher accuracies. Bottom graphs include this omitted set of accuracies.}
\label{fig:constantsqgl}
\end{figure}

Given that we look at 7 datasets, each of which belies a different result for how the presence of the synapse nonlinearity impact computation, it is difficult to draw a general conclusion from these results. Before considering the synapse condition, it is important to acknowledge that for the MNIST, FMNIST, KMNIST and EMNIST datasets, the binary tree outperformed the FCNN in the high multi-synaptic input repetition conditions and matched it in the low repetition conditions(Figure \ref{fig:synapse}A-D, Figure \ref{fig:sqgl_pval}A-D). We had assumed that the densely connected FCNN with a similar number of parameters as the sparsely connected $k$-tree would always have better performance. This result shows that this assumption is not the case. 
Also, if we are to simply compare the $k$-tree NaCaK conditions, then in MNIST and EMNIST datasets, the synapse nonlinearity architecture improves performance compared to the no-synapse architecture in the high-repetition conditions (Figure \ref{fig:synapse}A,D, Figure \ref{fig:sqgl_pval}A,D). The opposite may be seen in the FMNIST dataset (Figure \ref{fig:synapse}B, Figure \ref{fig:sqgl_pval}B), and it there is no difference in the KMNIST, SVHN, USPS, and CIFAR datasets (Figure \ref{fig:synapse}C,E-G, Figure \ref{fig:sqgl_pval}C,E-G). In the MNIST and KMNIST datasets, the $k$-tree performs better than the FCNN regardless of the presence of the synapse (Figure \ref{fig:synapse}A,C, Figure \ref{fig:sqgl_pval}A,C). In the FMNIST dataset, $k$-tree without synapse performs better than both FCNN conditions (Figure \ref{fig:synapse}B, Figure \ref{fig:sqgl_pval}B). From this variety of results, we can say that the combination of a binary tree constraint and a synaptic nonlinearity does not worsen model performance relative to all other conditions, and even has the potential to improve model performance above the supposed upper bound set by the FCNN model in higher-repetition cases.

\subsection*{A $k$-tree model with all biological constraints performs as well as or better than a model missing any one constraint}

In the binary tree model, the weight parameters connecting each node with downstream nodes can be argued to be analogous to axial conductances that is often found in multi-compartment models of neurons (Figure \ref{fig:model}D). Conductance is a scalar value, therefore its important to observe the impact of introducing the non-negative weight constraint in our specialized ANN model. In order to observe this constraint's impact on computation, we implement the $k$-tree with NaCaK nonlinearity architectures with or without synapse nonlinearities, and add the condition of constraining the model to have non-negative weights only. In Figure \ref{fig:constantsqgl} we compare all previous with or without synapse conditions to these new conditions.

We thus show that in all datasets, the non-negative $k$-tree without synapse performs very poorly in the low-multi-synaptic input repetition regime, or does not perform above our training failure accuracy threshold (0.55) at all (Figure \ref{fig:constantsqgl}A-G: Bottom, Figure \ref{fig:sqgl_pval}A-G). Importantly, this should be compared to the non-negative $k$-tree with synapses, which performs at a level comparable to other conditions (Figure \ref{fig:constantsqgl}, Figure \ref{fig:sqgl_pval}).
The non-negative $k$-tree with synapses performs at the level or better than of the positive and negative $k$-tree with synapses (Figure \ref{fig:constantsqgl}, Figure \ref{fig:sqgl_pval}).
In MNIST, KMNIST, and CIFAR10 conditions, the model with all constraints outperform FCNN with or without synapses (Figure \ref{fig:constantsqgl}A,C,G, Figure \ref{fig:sqgl_pval}A,C,G)
From this we can conclude that, using the $k$-tree NaCaK nonlinearity architecture, the combination of the presence of synapses and non-negative weights performs much better than the absence of synapses with non-negative weights. Overall, the most realistic $k$-tree model, with NaCaK nonlinearity, synapse nonlinearity and non-negative weights, performs at the level or better than a model missing any one of these constraints.

\section*{Experimental Procedures}

Much of this section can be derived from \cite{Jones2021}, and has been modified for the purpose of this study.

\subsection*{Computational Tasks}
Knowing that the output of a neuron is binary (presence or absence of an action potential), we chose to train our neuron model on a binary classification task. Using standard, high-dimensional, computer vision datasets, we used linear discriminant analysis (LDA) linear classifier to determine which 2 classes within each dataset were least linearly separable through training the LDA linear classifier and testing it on pairs of classes (Table \ref{tab-lda}). We used MNIST \citep{Lecun1998}, Fashion-MNIST \citep{Xiao2017}, EMNIST \citep{Cohen2017}, Kuzushiji-MNIST \citep{Clanuwat2018}, CIFAR-10 \citep{Krizhevsky2009}, Street View House Numbers (SVHN) \citep{Goodfellow2014}, and USPS \citep{Hastie2001} datasets.

\begin{table}[ht]
\tiny
\begin{center}
\begin{tabular}{ccccccc}
\toprule
\textbf{MNIST} & \textbf{FMNIST} & \textbf{EMNIST} & \textbf{KMNIST} & \textbf{CIFAR10} & \textbf{SVHN} & \textbf{USPS} \\ \hline
\\
0.8753 $\pm$   0.0120   & 0.6750 $\pm$ 0.0108   & 0.5821 $\pm$ 0.0180   & 0.6790 $\pm$ 0.0164  & 0.5254 $\pm$   0.0069     & 0.5186 $\pm$ 0.0102   & 0.8362 $\pm$ 0.0306 \\  \bottomrule
\end{tabular}
\end{center}

\caption{Linear Classifier Performance on Machine Learning Datasets}
\label{tab-lda}
\end{table}

\subsection*{Controls}
The control we use is a fully connected neural network (FCNN). The 2-layer FCNN is a comparable reference to see if $k$-tree performance meets or exceeds that of a densely connected network. The hidden layer of the FCNN is equal to twice the number of trees ($2k$) in the $k$-tree it is compared to and its output layer has 1 node.

\subsection*{Data Preprocessing}
We used datasets from the torchvision (version 0.5.0) python package. We then padded the 28 by 28 resolution images with zeros so that they were 32 x 32, and flattened the images to 1-D vectors. We then split the shuffled training set into training and validation sets (for MNIST, the ratio was 1:5 so as to let the validation set size match the test set), Then we split the resultant shuffled training set and validation set into 10 independent subsets. Each subset was used for a different cross-validation trial.

\subsection*{Model Architecture}
Using Pytorch (version 1.6.0), we designed the $k$-tree model architecture to be a feed forward neural network with sparse binary-tree connections. The weight matrices were implemented as simulated sparse tensors using the SparseLinear package created by rain-neuromorphics (updated August 18, 2020). Each node received 2 inputs and produces 1 output. To account for the sparsification, we altered the initialization of the weight matrices: we used standard “Kaiming normal” initialization with the gain of 1/density of the weight matrices. All nonzero weights were trained in model optimization. Lastly, the leak hyperparameter for LReLU was set to 0.01.

\subsection*{Model Training}
The model, inputs, and labels were loaded onto a Nvidia GeForce 1080 GPU using CUDA version 10.1. The batch size was 256. Early stopping was used such that after 60 epochs where no decrease in the validation loss is observed, training is stopped. We allowed the parameters of sparse linear layers of the $k$-tree and the synapse nonlinearity parameters to train. We did not allow the NaCaK conductance parameters ($\bar{g_{Na}}$, $\bar{g_{Ca}}$ and $\bar{g_{K}}$) to train. In all conditions, loss was calculated using mean binary cross entropy loss. In the $k$-tree condition with the synapse layer implemented, we also added a mean hinge loss when activations at any node were outside of the range of -70 and 50. We used the python library Hyperopt, which is based on the Tree-structured Parzen Estimator algorithm, in order to optimize the learning rates used for our Adam optimizer for each dataset and constraint condition. Each train-test loop was run for 10 trials with a different training subset each trial and the same test set every trial. Trial averages and standard errors were then calculated, and p-values were determined using two-tailed student's t-test (Figure \ref{fig:k-treenonlins_pval}-\ref{fig:sqgl_pval}).

\subsection*{Ion Channel Model Parameters}

The NaCaK function was built out of the IV curves of sodium \citep{Hodgkin1952}, calcium \citep{Miyasho2001}, and potassium \citep{Doiron2001} channels. The IV curves were derived from biophysical simulations of each ion channel using BRIAN2 \citep{Branco2010}, wherein the peak current was recorded for a range of voltage clamp settings. The parameters for these IV curves are depicted below.

The sodium channel \citep{Hodgkin1952} model had maximum conductance $\bar{g_{Na}} = 120 \text{mS}$ and a reversal potential $E_{Na} = 115 \text{mV}$.

\begin{align}
I_{m} & = g_{Na}m^3h(V - E_{Na}) \\ \nonumber
\alpha_m & = \frac{1}{\exp(\frac{25\text{mV} - V}{10\text{mV}})} \\ \nonumber
\beta_m & = 4\exp(\frac{-V}{18\text{mV})} \\ \nonumber
\alpha_h & = 0.07\exp(\frac{-V}{20\text{mV}}) \\ \nonumber
\beta_h & = \frac{1}{1 + \exp(\frac{30\text{mV} - V}{10\text{mV})}}.
\end{align}

The calcium channel \citep{Miyasho2001} model had maximum conductance $\bar{g_{Ca}} = 100 \text{mS}$ and a reversal potential $E_{Ca} = 135 \text{mV}$.

\begin{align}
I_{m} & = g_{Ca}mh(V - E_{Ca}) \\ \nonumber
\alpha_m & = \frac{2.6}{1 + \exp(\frac{V + 21\text{mV}}{-8\text{mV}})} \\ \nonumber
\beta_m & = \frac{0.18}{1 + \exp(\frac{V + 21\text{mV}}{-8\text{mV}})} \\ \nonumber
\alpha_h & = \frac{0.0025}{1 + \exp(\frac{V + 40\text{mV}}{8\text{mV}})} \\ \nonumber
\beta_h & = \frac{0.19}{1 + \exp(\frac{V + 50\text{mV}}{-10\text{mV}})}. \\ \nonumber
\end{align}

The potassium channel \citep{Doiron2001} model had maximum conductance $\bar{g_{K}} = 100 \text{mS}$ and a reversal potential $E_K = -88.5$.

\begin{align}
I_{m} & = g_{K}m^3h(V - E_{K}) \\ \nonumber
m_{\infty} & = \frac{1}{1 + \exp(\frac{V}{-19\text{mV}})} \\ \nonumber
\tau_m & = 0.8 \\ \nonumber
h_{\infty} & = \frac{1}{1 + \exp(\frac{V + 3\text{mV}}{40\text{mV}})} \\ \nonumber
\tau_h & = 1.5. \\ \nonumber
\end{align}

\section*{Discussion}

Here we introduced three important biological constraints to models of hierarchical dendritic computation. For each of the constraints, we asked how it impacted the model's computational performance on several machine learning tasks. We found that the NaCaK nonlinearity performs as well if not better than the commonly used deep learning activation functions ReLU, LReLU, and sigmoid nonlinearities. Implementing synapse nonlinearities while using the NaCaK nonlinearity does not worsen binary tree model performance. When using the this nonlinearity, constraining the weight parameters to be non-negative can greatly penalize performance in low multi-synaptic input repetition regimes, but implementing synapses along with this constraint rescues model performance. The combination of all three constraints makes a model that performs as well as if not better than models without constraints. These constraints capture many limitations of real dendrites, suggesting that neurons with real dendrites may be able to actually solve the kinds of complex machine learning problems used in the deep learning field.

\subsection*{Limitations}

Our dendritic binary tree model is technically a feed-forward ANN, which is an abstraction away from the temporally dynamic nature of neurons \citep{Hines1997}. In order to allow our model to work within this feed-forward context, we used a steady-state synaptic voltage nonlinearity approximation instead of real temporal conductance dynamics. Important neuronal phenomena, like backpropagating action potentials or dendritic direction selectivity \citep{London2005} cannot be reproduced using this model. In addition, inputs to neurons have a temporal dimension as well, so the input we use for this feed-forward model is not biologically plausible. An addition of real recurrent dynamics could render our model similar to those often implemented in compartmental modeling \citep{Hines1997,Bower2013}. While it would make conceptualization harder, implementing recurrent connections and using input with a time dimension could produce a more biologically plausible model and could be a very interesting constraint to add to the model.

The voltage-gated ion channel derived function, the NaCaK function, is a linear weighted sum of sodium, calcium, and potassium I-V curves where the weights of the sum are analogous to maximal conductance, which is a function of ion channel density. A limitation of this conception of dendritic nonlinearity is that we chose only 3 out of many ion channels that are present in a dendritic tree of any one neuron. We chose representative  sodium \citep{Hodgkin1952}, potassium \citep{Doiron2001}, and calcium \citep{Miyasho2001} channels. It is an open question which ion channels would be best for making a general dendrite nonlinearity for the purposes of a model like the one we used. We also did not try to model any specific neuron type, so it would be very interesting to ask how distinct neuron types could support the computations we are analyzing. Follow-up work could be more specific about the kinds of neuron model they choose to constrain the model to. 

Another limitation of this conception of dendritic nonlinearity is that the ratio of these ion channel density weights is, unrealistically, equal. Computational studies \citep{Huys2006,Hay2011} have found that that not only is there no even ratio of these ion channels, this ratio is most certainly not consistent across the entirety of the morphology of the dendritic tree. Future work that allows these weights to be optimized might be fruitful for possibly explaining ion channel density distributions in intricate dendritic morphologies. A combination of more realistic morphologies than the $k$-tree and a learnable NaCaK function for each node might introduce more biologically relevant degrees of freedom that could impact model computational performance.

High count multi-synaptic input repetitions to the terminals of separate dendritic sub-trees is a theoretical extreme of what real neurons are doing. These theoretical multi-synaptic boutons (MSB) assume that the repeated synaptic inputs go to different dendritic sub-trees. The literature shows support of same-branch MSBs and different-branch MSBs \citep{Kincaid1998, Jones1997}, but it is unclear if different-branch MSBs are to dendritic branches from completely different dendritic sub-trees. In addition, much evidence shows that dendritic trees are highly asymmetric \citep{RamonyCajal1894}, with synapses being at terminal nodes along lengths of dendritic node chains. It has been shown that these asymmetric trees are difficult to train on ML tasks using LReLU nonlinearities \citep{Jones2021}. Perhaps this work can inspire further work returning to this asymmetric tree architecture using the biological constraints we used in order to see if that allows the architecture to train better. In the context of the symmetric binary tree we use with multiple sub-trees, it may be more biologically relevant to look at performances of the low and medium repetition conditions.

The input-output functions from the deep learning literature that we use here to quantify computational capacity are a bad approximation to the problems solved by real neurons. The literal input-output (I/O) function of mapping an input of a 1-D, complex pixelated image to the presence or absence of an action potential is  clearly biologically unrealistic. For example, in the visual system, the outputs of LGN neurons would have different structure \citep{DiCarlo2012}. However, we use this simple I/O function to explore the computational possibilities of an individual neuron. It would be interesting to see if the real functions computed by neurons are easier or harder.

In this study we used the backpropagation of error learning algorithm. Real neurons may learn this way or in entirely different ways and this is an active and controversial area of study \citep{Lillicrap2020}. A critique of the backpropagation of error algorithm is that it remains unobserved how the neuron achieves error feedback to the synapses, allowing the synapses to change based off of a gradient signal \citep{Lillicrap2020}. Alternatives to backpropagation of error, such as feedback alignment \citep{Lillicrap2016}, predictive coding \citep{Millidge2020}, or equilibrium propagation \cite{Scellier2017}, could be used in this testing context. Further work may also show how plasticity rules in combination with third-factor signaling \citep{Richards2019a} could possibly provide the error feedback a single neuron needs to adjust its synaptic weights. As for plasticity of dendritic branches, they are most plastic during early neuron development \citep{Koleske2013}. Work may be needed to make models that distinguish between fully plastic neurons during early development and mature neurons which only have plastic synapses. Further work like this on neuron model learnability is needed in order to evaluate potential learning models in the future.

It is possible now to directly study dendritic integration in real neurons, although it is hard to do justice to their high-dimensional inputs \citep{Spruston2009}. One can record electrophysiologically  dendrite-to-soma firing I/O functions that have the specific constraint of a multi-dendritic input corresponding to an action potential output. Further work using 2-photon microscopy, glutamate uncaging, dendritic voltage dyes, and electrophysiological recording could be useful in generating the kind of I/O function data that can produce further constraints on dendritic computation. It would be very interesting to calibrate neurons against physiological data and then ask about their computational limitations.

\subsection*{Contributions}

We found that the NaCaK nonlinearity and swish nonlinearity performed the best in the sparse binary tree architecture. The NaCaK nonlinearity relates to the transfer function literature that has shown that smooth functions that are locally similar to ReLU tend to perform well \citep{Xu2015}. We also point out that the local properties of the neuron share some aspects of ResNets \citep{Xie2017, He2016}. Without local nonlinearities, a node simply copies its parents, producing the unity function. ResNets have the capacity to have better performances in machine learning benchmark tasks than a traditional multi-layer perceptron \citep{Xie2017}. Also, the nonlinear component of the NaCaK function contributes gradients (Figure \ref{fig:derivatives}) used by the backpropagation of error algorithm used to train the model, allowing the neuron model to learn the binary classification task more effectively than a completely linear model (See and compare performances to Table \ref{tab-lda}). These components of the NaCaK function show that producing learning systems that are more similar to those of real neurons may benefit from our considerations, e.g. for the design of specialized hardware. Interestingly, the sigmoid nonlinearity generally did worse than the NaCaK nonlinearity, suggesting that the brain has evolved a good transfer function.

The three constraints we implemented either did not worsen or improved the performance of our dendritic binary tree model in mid-to-high multi-synaptic repetition regimes. In low repetition regimes, binary trees using NaCaK nonlinearity and the non-negative weight constraint suffered in performance accuracy without the synapse nonlinearity. We suggest two potential reasons for why the synapse nonlinearity rescued performance. First, the synapse nonlinearity significantly expands the model size by introducing 4 more parameters per input. The synapse nonlinearity merely represents a few degrees of freedom that exist in a real synapse. \citep{Koch1999} This increase in synapse complexity in contrast to the more simple traditional way of representing synaptic strength with a scalar could be a reason why performance is rescued in this case. Second, it is important to note that the synapse nonlinearity maps the input, which ranges from 0 to 1, to synaptic voltage activation in millivolts, and the change in range of the NaCaK nonlinearity is across a wide domain between -70 and 50 mV (Figures \ref{fig:derivatives}, \ref{fig:syn_act_heat}, and \ref{fig:syn_act_hist} ). Without the synapse nonlinearity, the activation in the binary tree would likely only sit close to 0. Having both the range of activations between -70 and 50 that the synapse nonlinearity enforces and the NaCaK nonlinearity with a matching range likely takes full advantage of the nonlinearities as activation travels down the binary tree. Though it is still unclear why non-negative weight constrained models need this matching in order to perform well, this may be why the combination of all three constraints leads to consistent or better performance.

Using the biological constraints we introduced a variety of free parameters analogous to properties in real neurons. Upon further analysis we found that in a model trained on MNIST with all three biological constraints the synaptic axial conductance weights (Figure \ref{fig:syn_act_hist}) were close to the ranges found in \cite{Araya2014}. The dendritic axial conductance weights (Figure \ref{fig:dendrite_activations}A) were similar in magnitude. The resultant activations were also within the expected range of between -70 and 50 mV (Figures \ref{fig:syn_act_hist} and \ref{fig:dendrite_activations}B,C). These results convince us that the model, when given biological constraints, yields biologically relevant behavior.

We take the time here to discuss the term "constraint." Given that our formulation of the synapse nonlinearity technically expands the size of our model, it may be seen as inappropriate to call the synapse nonlinearity a constraint. Interestingly, the machine learning field will tend to view the number of parameters as main constraint \citep{scholkopf2002} while the neuroscience field may rather view the existence of the channels that exist with their biological properties as a constraint - after all there are simply computations that are impossible under the biological constraints (e.g. those where the membrane voltage is >100mV). As such, we do not agree with just viewing the number of parameters as a constraint. Before constructing a synapse in a model and before knowing how that construction impacts model performance, we would be agnostic about whether the property limits or enhances model performance. In addition, we call this property a constraint because of its grounding in observed reality that abstracted models do not include or engage with. Therefore we call the synaptic nonlinearity a constraint in reference to it grounding or "constraining" our model to biological realism.

We introduced three biological constraints to an abstracted computational model of a neuron with dendrites, and observed the impacts of these constraints on computational performance on well-defined ML tasks. We found that NaCaK nonlinearity which approximates real dendrite physiology clearly outperforms other nonlinearities. The addition of a synapse nonlinearity and non-negative weight constraint does not worsen performance on ML tasks, and in some cases improves performance. This more biologically plausible model can be trained on well-defined tasks, which opens further possibilities of exploring what neurons are capable of computationally. Our findings contribute to the theoretical evidence that dendritic computation can approximate complex nonlinear I/O functions, even with the constraints imposed by the components of dendrites.

\section*{Acknowledgments}

We would like to acknowledge the members of the Kording Lab, specifically Roozbeh Farhoodi, Ben Baker and Ari Benjamin, for help in the development of this project. This work was funded by grants from the National Institute of Health, National Science Foundation, and Howard Hughes Medical Institute.

\section*{Code}
The code for this project can be found at the following github repository:

https://github.com/ilennaj/ktree$\_$constraints

\newpage

\bibliography{bibliography}

\newpage
\section*{Supplementary Figures}
\setcounter{figure}{0}
\renewcommand{\thefigure}{S\arabic{figure}}

\begin{figure}[ht]
\centering
\includegraphics[width=0.8\linewidth]{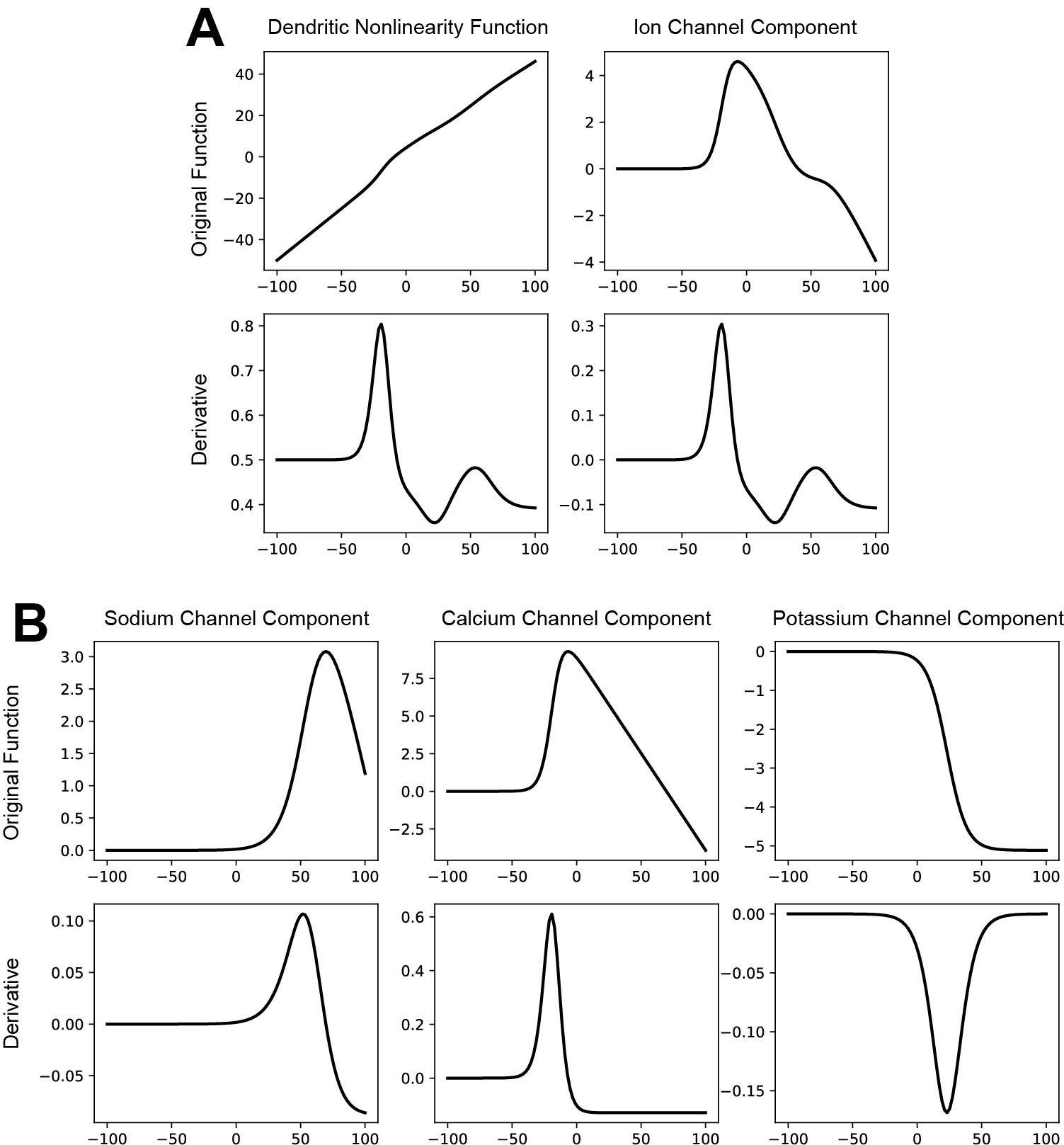}
\caption{Dendritic Nonlinearity implemented as a linear combination of Sodium, Calcium, and Potassium Current-Voltage Curves. Dendritic Nonlinearity is called the NaCaK function. A) Top: Original NaCaK Function and its Ion channel component. Bottom: Derivatives of these functions. B) Top: Sodium, Calcium, and Potassium channel fitted I-V curves. Bottom: Derivatives of these I-V curves.}
\label{fig:derivatives}
\end{figure}

\begin{figure}[ht]
\centering
\includegraphics[width=0.8\linewidth]{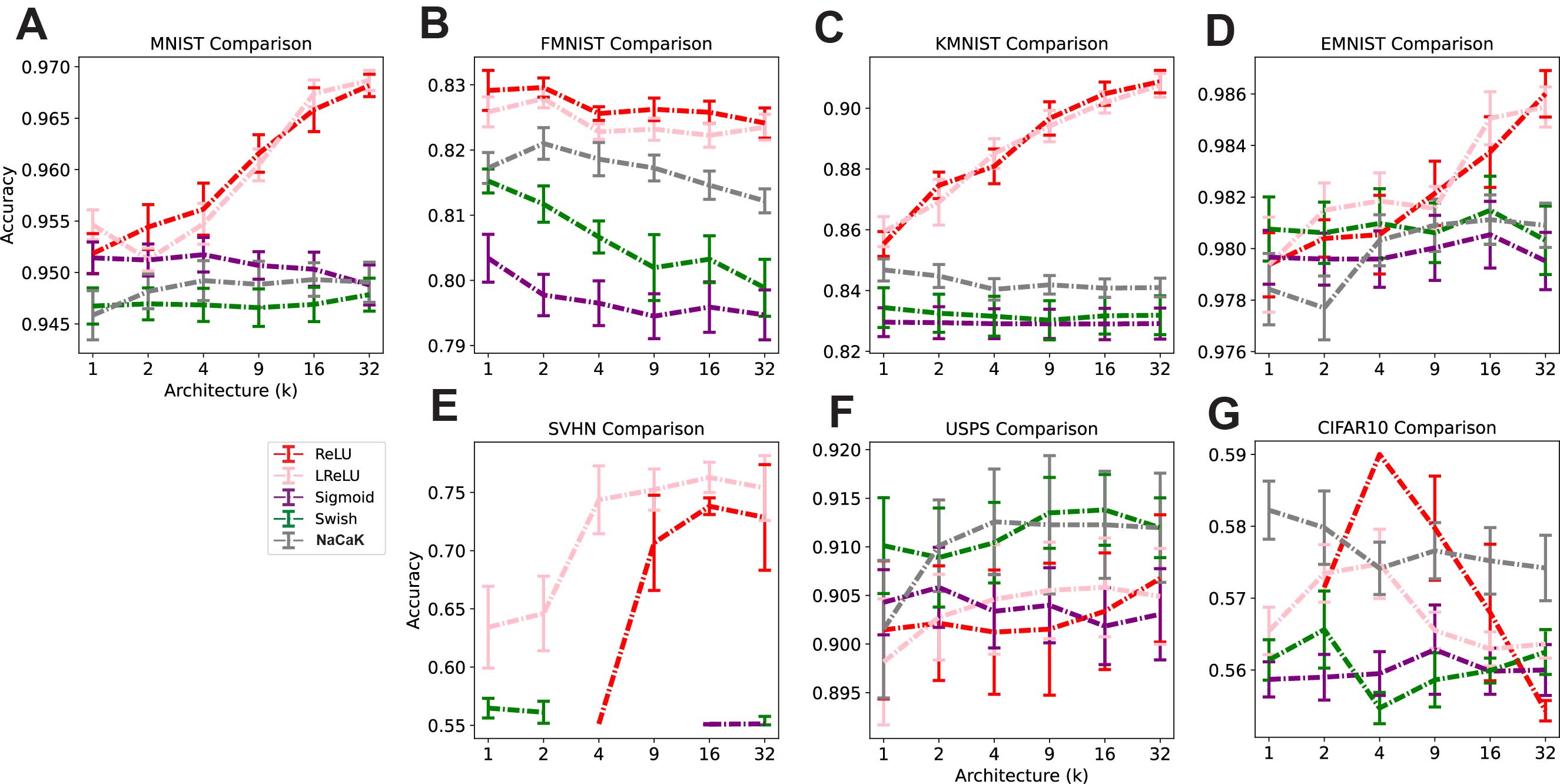}
\caption{Comparing Nonlinear Activation Functions for the Dendrite Nodes in the FCNN For Each Dataset. Architecture ranges from low hidden layer width (k=1) to high hidden layer width (k=32) and mean accuracy of each model is displayed with standard error bars. Number of trials ranges from 1-10, and trial accuracy was omitted if the training was deemed to fail (accuracy threshold 0.55). Table \ref{tab-fcnn_nonlins}  lists trial counts. Red: ReLU, Pink: LReLU, Purple: Sigmoid, Green: Swish, Grey: NaCaK Nonlinearity}
\label{fig:fcnn_nonlins}
\end{figure}

\begin{figure}[ht]
\centering
\includegraphics[width=0.7\linewidth]{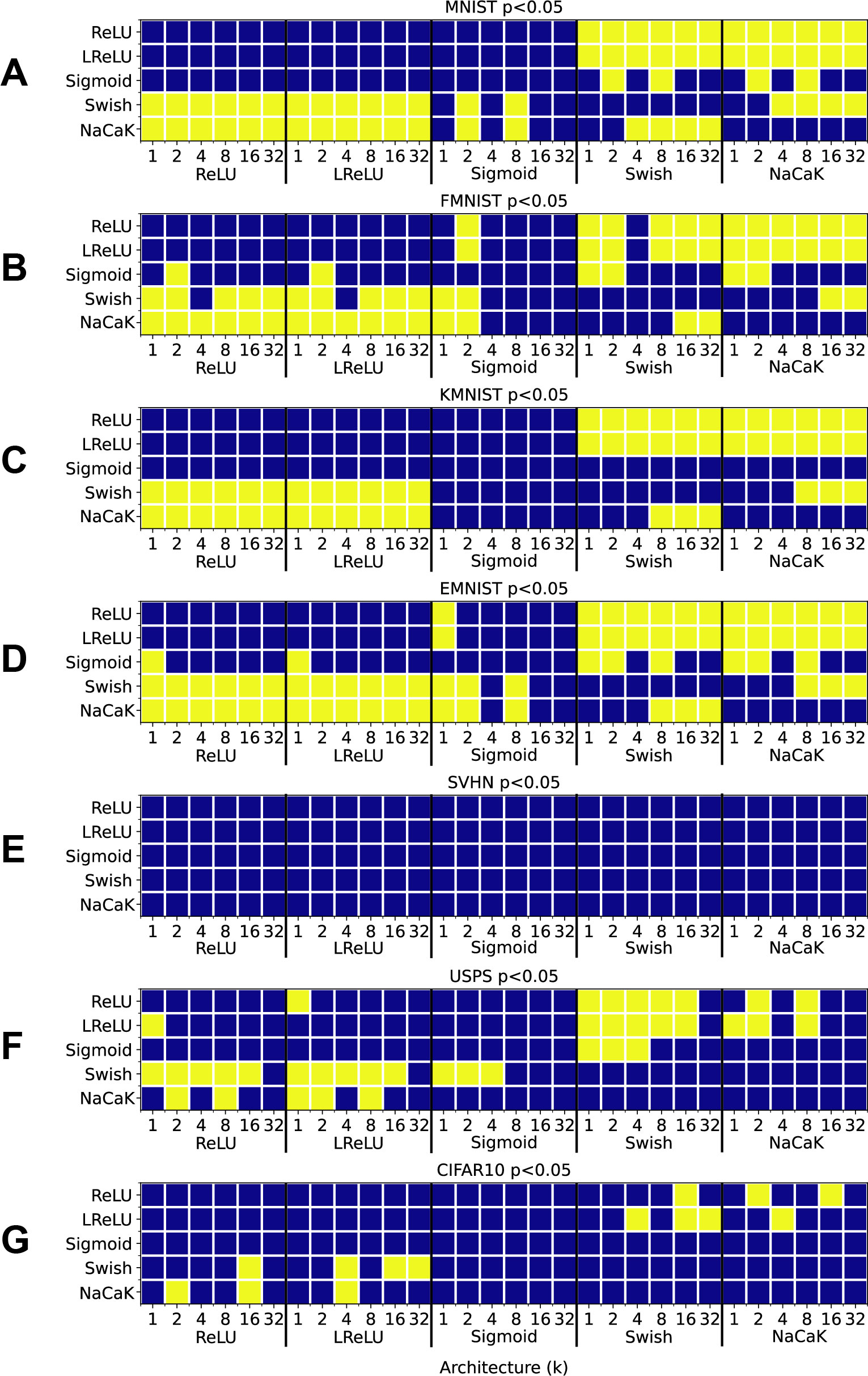}
\caption{Significance Indicator Matrix for Comparing Nonlinear Activation Functions for the Dendrite Nodes in the k-tree. Yellow squares indicate p<0.05 significance and blue squares indicate p>0.05 significance. Architecture ranges from low multi-synaptic input repetitions (k=1) to high repetitions (k=32). Number of trials ranges from 1-10, and if there were not enough trials to run a t-test, then the block is automatically blue.}
\label{fig:k-treenonlins_pval}
\end{figure}

\begin{figure}[ht]
\centering
\includegraphics[width=0.7\linewidth]{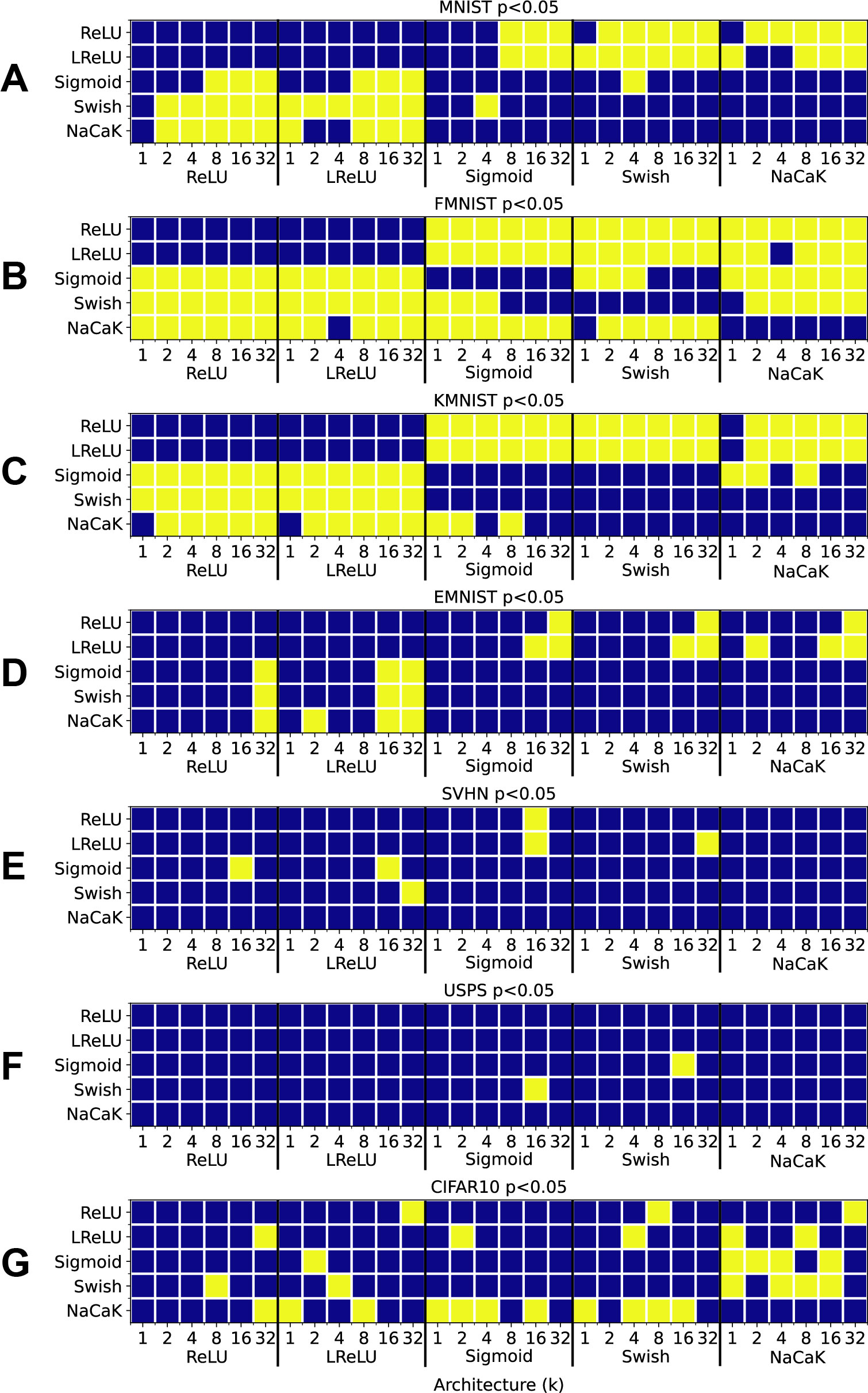}
\caption{Significance Indicator Matrix for Comparing Nonlinear Activation Functions for the Dendrite Nodes in the FCNN. Yellow squares indicate p<0.05 significance and blue squares indicate p>0.05 significance. Architecture ranges from low hidden layer width (k=1) to high hidden layer width (k=32). Number of trials ranges from 1-10, and if there were not enough trials to run a t-test, then the block is automatically blue.}
\label{fig:fcnn_nonlins_pval}
\end{figure}

\begin{figure}[ht]
\centering
\includegraphics[width=0.7\linewidth]{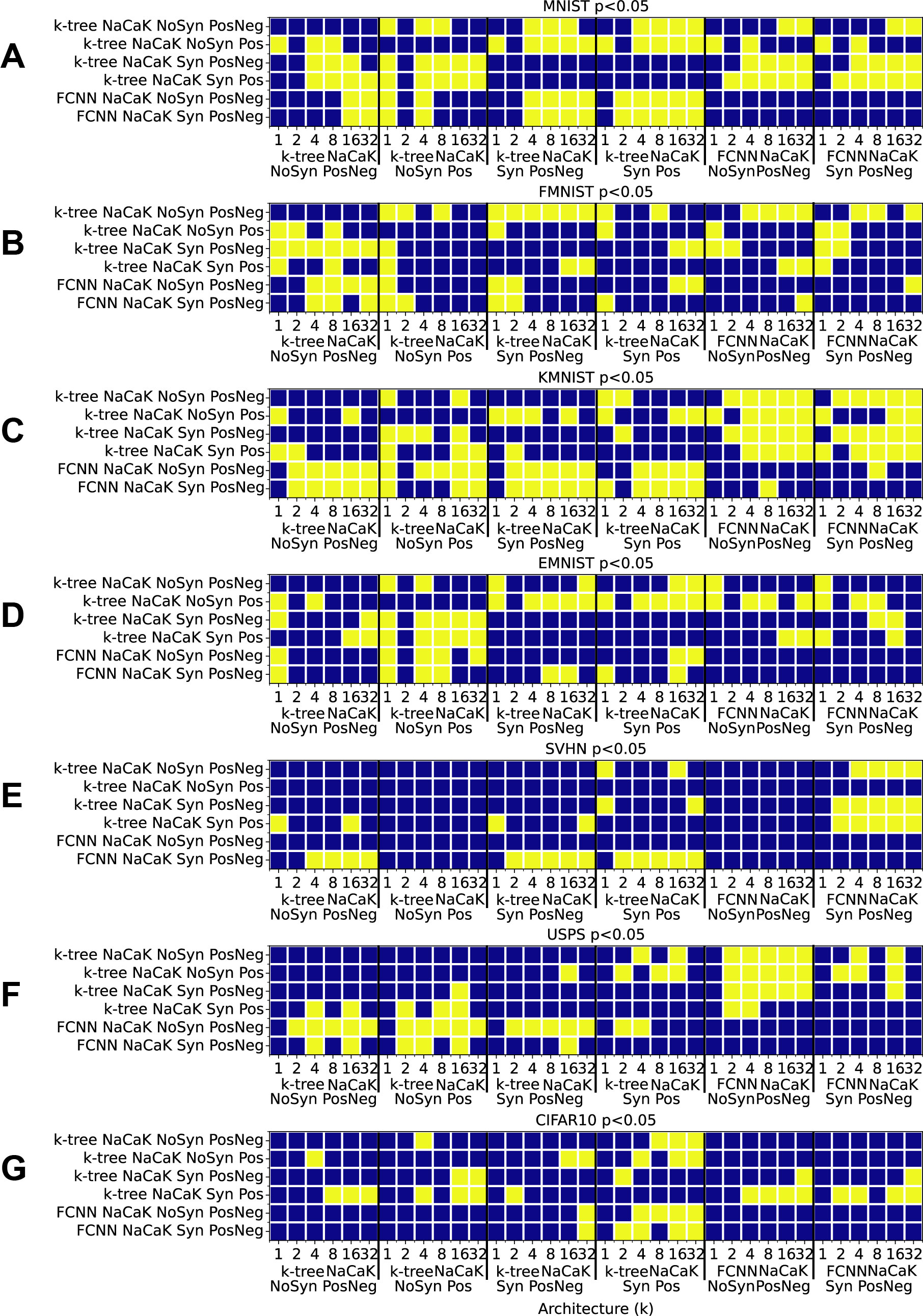}
\caption{Significance Indicator Matrix for Comparing $k$-tree and FCNN Models With or Without Synapse Nonlinearity or Non-negativity Weight Constraint. Yellow squares indicate p<0.05 significance and blue squares indicate p>0.05 significance. Architecture ranges from low multi-synaptic input repetitions (k=1) to high repetitions (k=32). Number of trials ranges from 1-10, and if there were not enough trials to run a t-test, then the block is automatically blue. NoSyn = No Synapse Nonlinearity. Syn = Synapse Nonlinearity. PosNeg = Positive and Negative Weights. Pos = Non-negative Weights.}
\label{fig:sqgl_pval}
\end{figure}

\begin{figure}[ht]
\centering
\includegraphics[width=0.8\linewidth]{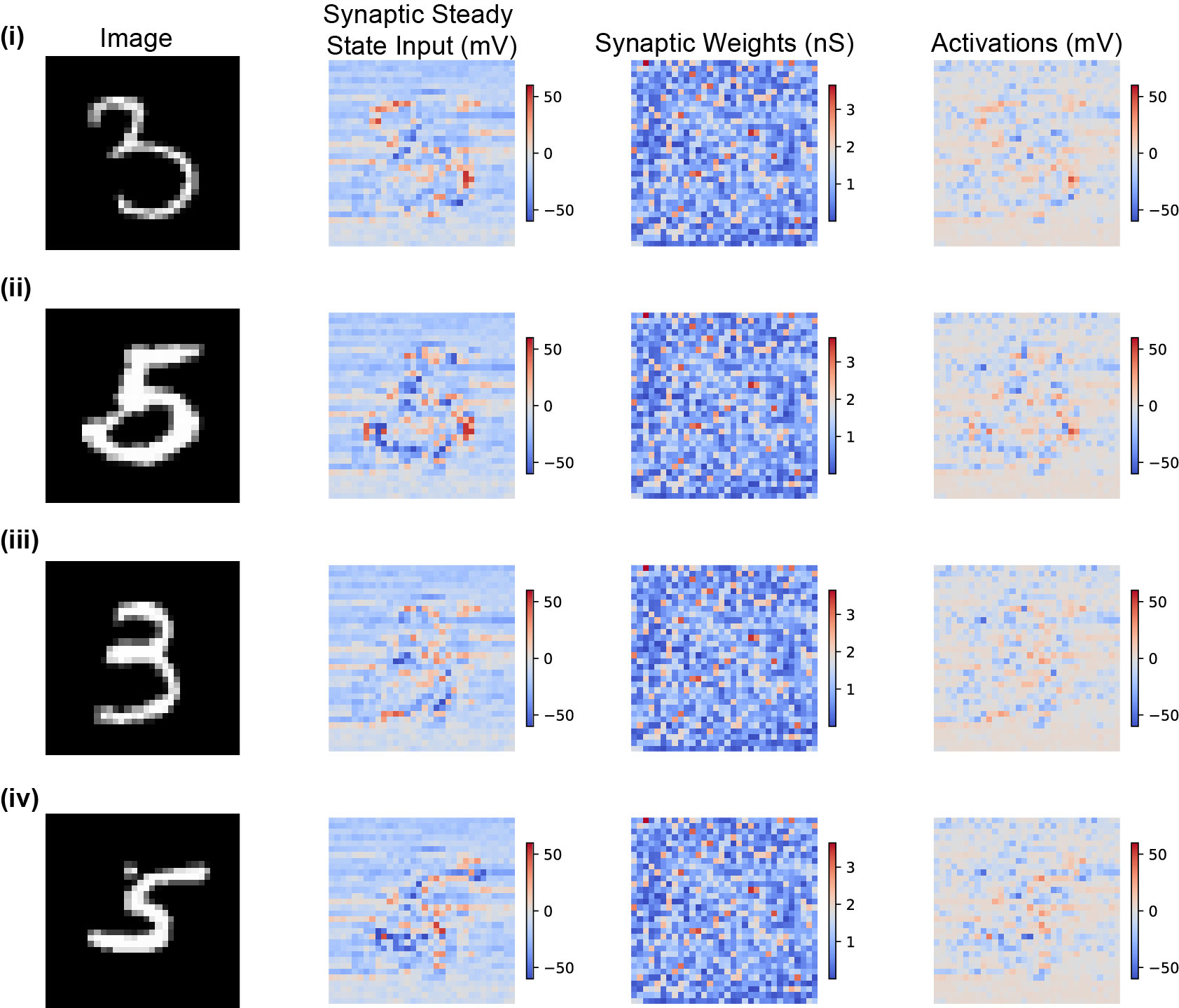}
\caption{Heatmaps of Synaptic Activations. The images (left) are vectorized and input into a synaptic nonlinearity layer, which yields a synaptic steady state input between -70 and 50 mV. The product of the synaptic weights, analogous to synaptic bouton neck conductances, and the synaptic steady state inputs then lead a final synaptic activation (right) that is received by downstream dendrite nodes.}
\label{fig:syn_act_heat}
\end{figure}

\begin{figure}[ht]
\centering
\includegraphics[width=0.8\linewidth]{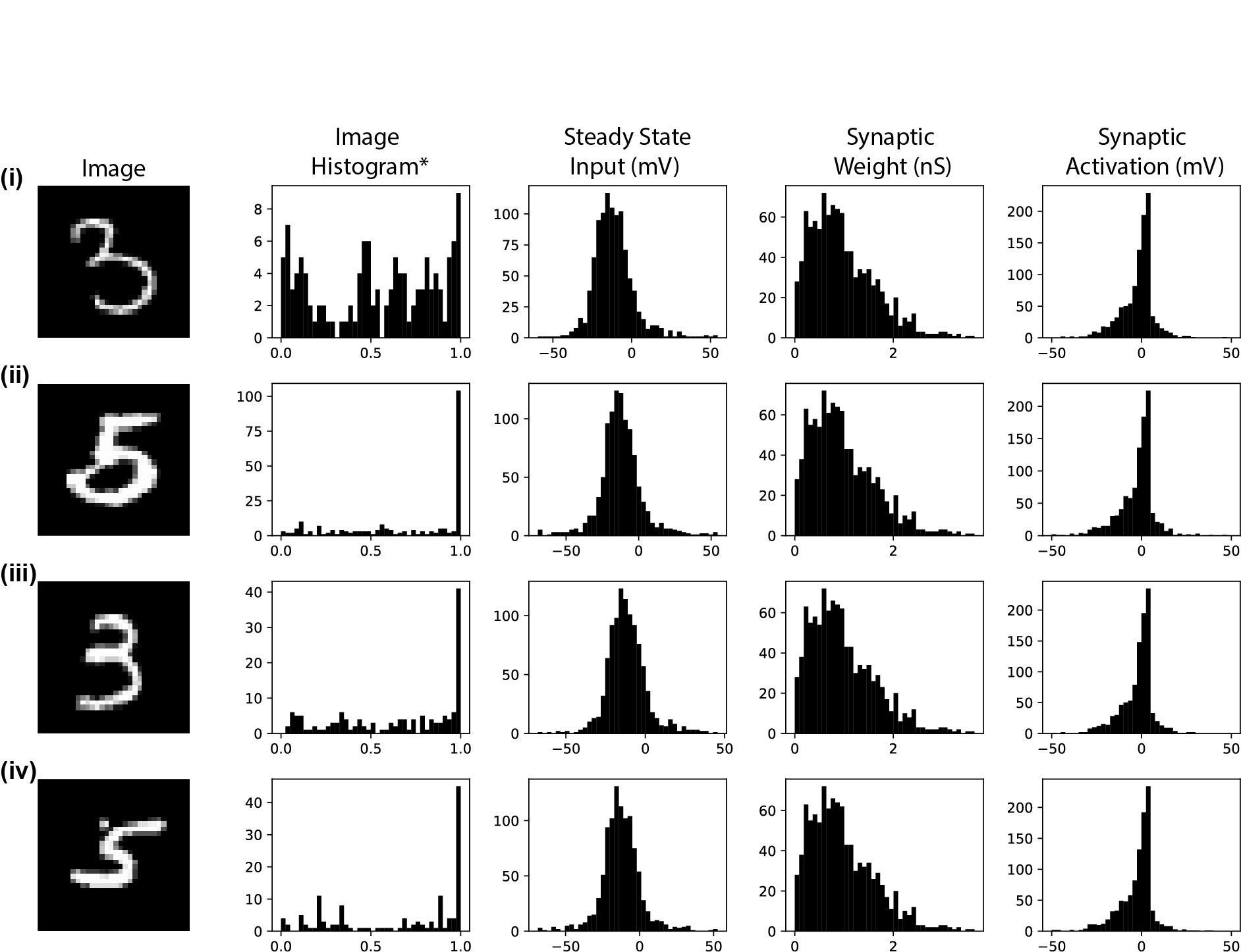}
\caption{Histograms of Synaptic Activations. The images (left) are vectorized and input into a synaptic nonlinearity layer, which yields a synaptic steady state input between -70 and 50 mV. The product of the synaptic weights, analogous to synaptic bouton neck conductances, and the synaptic steady state inputs then lead a final synaptic activation (right) that is received by downstream dendrite nodes.}
\label{fig:syn_act_hist}
\end{figure}

\begin{figure}[ht]
\centering
\includegraphics[width=0.8\linewidth]{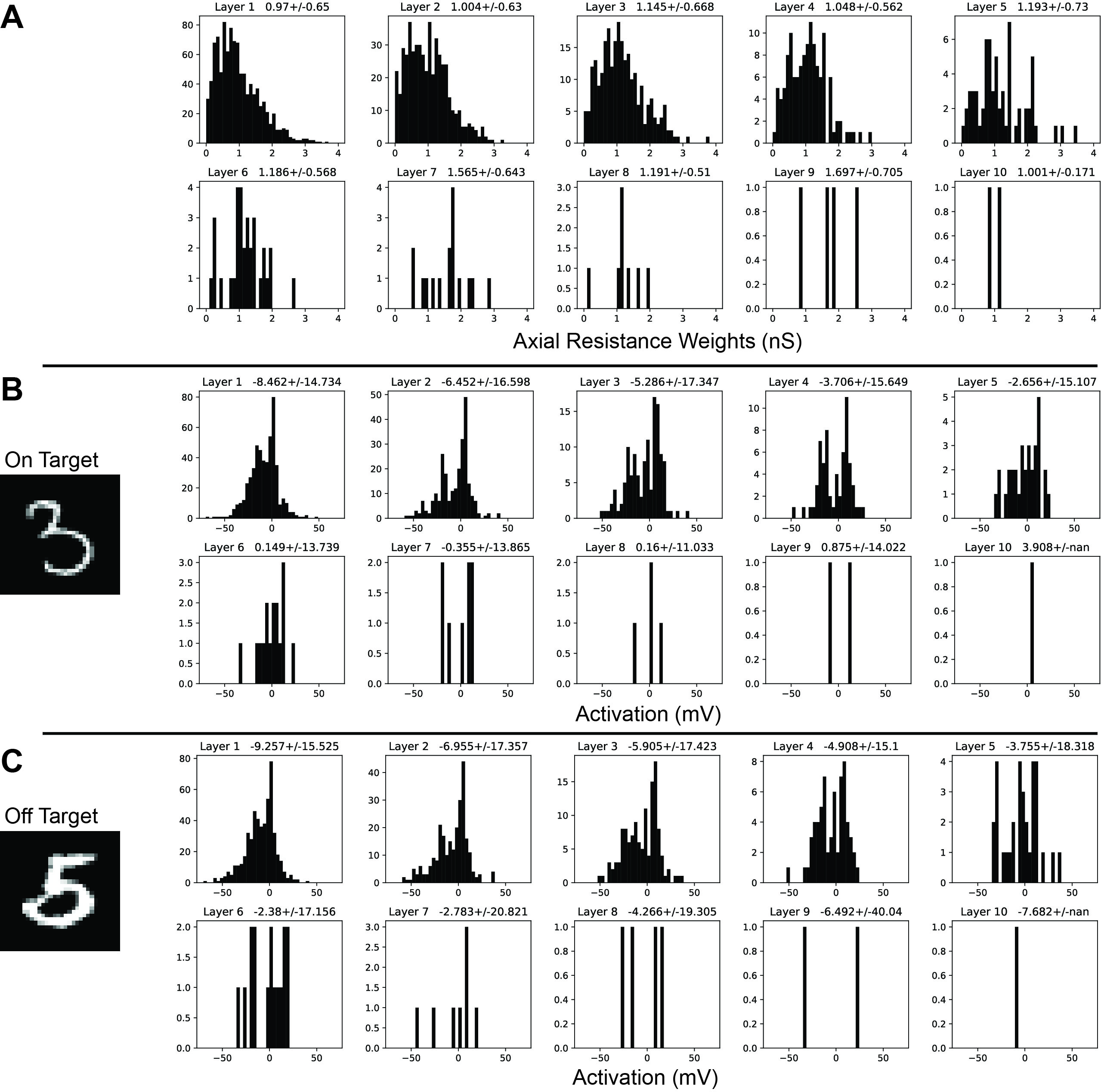}
\caption{Distributions of Dendritic Node Weights and Activations of a Trained 1-tree. A) Dendritic node weight distribution (in nanoSiemens) in layers 1-10 of a 1-tree. Layers are labeled with layer number, mean and standard deviation. B and C) Dendritic node activation distribution (in milliVolts) in layers 1-10 of a 1-tree for "On Target" (B) and "Off Target (C) stimuli. Layers are labeled with layer number, mean, and standard deviation.}
\label{fig:dendrite_activations}
\end{figure}

\begin{table}[ht]
\tiny
\begin{center}
\begin{tabular}{lrrrrrr}
\toprule
&&&MNIST \\
\midrule
      & 1-tree & 2-tree & 4-tree & 8-tree & 16-tree & 32-tree \\
\midrule
ReLU &      8 &      9 &     10 &      9 &      10 &      10 \\
LReLU &      8 &     10 &     10 &     10 &      10 &      10 \\
Sigmoid &      1 &      5 &      0 &      2 &       0 &       0 \\
Swish &     10 &     10 &     10 &     10 &      10 &      10 \\
NaCaK &     10 &     10 &     10 &     10 &      10 &      10 \\
\bottomrule
\end{tabular}

\begin{tabular}{lrrrrrr}
\toprule
&&&FMNIST \\
\midrule
      & 1-tree & 2-tree & 4-tree & 8-tree & 16-tree & 32-tree \\
\midrule
ReLU &      7 &     10 &      8 &     10 &      10 &      10 \\
LReLU &      7 &     10 &     10 &     10 &      10 &      10 \\
Sigmoid &      2 &      3 &      0 &      0 &       0 &       1 \\
Swish &      4 &      7 &      8 &      9 &      10 &      10 \\
NaCaK &      9 &     10 &     10 &     10 &      10 &      10 \\
\bottomrule
\end{tabular}

\begin{tabular}{lrrrrrr}
\toprule
&&&KMNIST \\
\midrule
      & 1-tree & 2-tree & 4-tree & 8-tree & 16-tree & 32-tree \\
\midrule
ReLU &      8 &      9 &      9 &     10 &      10 &      10 \\
LReLU &      7 &      7 &     10 &     10 &      10 &      10 \\
Sigmoid &      1 &      1 &      1 &      0 &       0 &       0 \\
Swish &     10 &     10 &     10 &     10 &      10 &      10 \\
NaCaK &      9 &     10 &     10 &     10 &      10 &      10 \\
\bottomrule
\end{tabular}

\begin{tabular}{lrrrrrr}
\toprule
&&&EMNIST \\
\midrule
      & 1-tree & 2-tree & 4-tree & 8-tree & 16-tree & 32-tree \\
\midrule
ReLU &     10 &     10 &     10 &     10 &      10 &      10 \\
LReLU &     10 &     10 &     10 &     10 &      10 &      10 \\
Sigmoid &      8 &      3 &      1 &      2 &       0 &       0 \\
Swish &     10 &     10 &     10 &     10 &      10 &      10 \\
NaCaK &     10 &     10 &     10 &     10 &      10 &      10 \\
\bottomrule
\end{tabular}

\begin{tabular}{lrrrrrr}
\toprule
&&&SVHN \\
\midrule
      & 1-tree & 2-tree & 4-tree & 8-tree & 16-tree & 32-tree \\
\midrule
ReLU &      0 &      0 &      0 &      1 &       0 &       1 \\
LReLU &      1 &      1 &      1 &      1 &       1 &       4 \\
Sigmoid &      0 &      0 &      0 &      0 &       0 &       0 \\
Swish &      4 &      2 &      2 &      3 &       3 &       3 \\
NaCaK &      2 &      3 &      4 &      4 &       3 &       2 \\
\bottomrule
\end{tabular}

\begin{tabular}{lrrrrrr}
\toprule
&&&USPS \\
\midrule
      & 1-tree & 2-tree & 4-tree & 8-tree & 16-tree & 32-tree \\
\midrule
ReLU &      6 &      8 &     10 &     10 &      10 &      10 \\
LReLU &      6 &      9 &     10 &     10 &      10 &      10 \\
Sigmoid &      3 &      3 &      2 &      0 &       0 &       0 \\
Swish &     10 &     10 &     10 &     10 &      10 &      10 \\
NaCaK &      8 &      6 &      7 &     10 &      10 &      10 \\
\bottomrule
\end{tabular}

\begin{tabular}{lrrrrrr}
\toprule
&&&CIFAR10 \\
\midrule
      & 1-tree & 2-tree & 4-tree & 8-tree & 16-tree & 32-tree \\
\midrule
ReLU &      2 &      4 &      4 &      5 &       5 &       6 \\
LReLU &      2 &      1 &      5 &      3 &       7 &       6 \\
Sigmoid &      0 &      0 &      0 &      0 &       0 &       0 \\
Swish &      1 &      4 &      3 &      2 &       4 &       6 \\
NaCaK &     10 &      9 &      8 &      8 &       9 &       8 \\
\bottomrule
\end{tabular}
\end{center}
\caption{Trial Numbers for Figure \ref{fig:k-treenonlins}, Comparing Nonlinear Activation Functions for the Dendrite Nodes in the $k$-tree For Each Dataset.}
\label{tab-ktree_nonlins}
\end{table}

\begin{table}[ht]
\tiny
\begin{center}
\begin{tabular}{lrrrrrr}
\toprule
&&MNIST \\
\midrule
                     & 1-tree & 2-tree & 4-tree & 8-tree & 16-tree & 32-tree \\
\midrule
k-tree SQGL NoSyn PosNeg &     10 &     10 &     10 &     10 &      10 &      10 \\
k-tree SQGL Syn PosNeg &      9 &     10 &     10 &     10 &      10 &      10 \\
FCNN SQGL NoSyn PosNeg &     10 &     10 &     10 &     10 &      10 &      10 \\
FCNN SQGL Syn PosNeg &     10 &     10 &     10 &     10 &      10 &      10 \\
\bottomrule
\end{tabular}

\begin{tabular}{lrrrrrr}
\toprule
&&FMNIST \\
\midrule
                     & 1-tree & 2-tree & 4-tree & 8-tree & 16-tree & 32-tree \\
\midrule
k-tree SQGL NoSyn PosNeg &      9 &     10 &     10 &     10 &      10 &      10 \\
k-tree SQGL Syn PosNeg &     10 &     10 &     10 &     10 &      10 &      10 \\
FCNN SQGL NoSyn PosNeg &     10 &     10 &     10 &     10 &      10 &      10 \\
FCNN SQGL Syn PosNeg &     10 &     10 &     10 &     10 &      10 &      10 \\
\bottomrule
\end{tabular}

\begin{tabular}{lrrrrrr}
\toprule
&&KMNIST \\
\midrule
                     & 1-tree & 2-tree & 4-tree & 8-tree & 16-tree & 32-tree \\
\midrule
k-tree SQGL NoSyn PosNeg &      9 &     10 &     10 &     10 &      10 &      10 \\
k-tree SQGL Syn PosNeg &     10 &     10 &     10 &     10 &      10 &      10 \\
FCNN SQGL NoSyn PosNeg &     10 &     10 &     10 &     10 &      10 &      10 \\
FCNN SQGL Syn PosNeg &     10 &     10 &     10 &     10 &      10 &      10 \\
\bottomrule
\end{tabular}

\begin{tabular}{lrrrrrr}
\toprule
&&EMNIST \\
\midrule
                     & 1-tree & 2-tree & 4-tree & 8-tree & 16-tree & 32-tree \\
\midrule
k-tree SQGL NoSyn PosNeg &     10 &     10 &     10 &     10 &      10 &      10 \\
k-tree SQGL Syn PosNeg &     10 &     10 &     10 &     10 &      10 &      10 \\
FCNN SQGL NoSyn PosNeg &     10 &     10 &     10 &     10 &      10 &      10 \\
FCNN SQGL Syn PosNeg &     10 &     10 &     10 &     10 &      10 &      10 \\
\bottomrule
\end{tabular}

\begin{tabular}{lrrrrrr}
\toprule
&&SVHN \\
\midrule
                     & 1-tree & 2-tree & 4-tree & 8-tree & 16-tree & 32-tree \\
\midrule
k-tree SQGL NoSyn PosNeg &      2 &      3 &      4 &      4 &       3 &       2 \\
k-tree SQGL Syn PosNeg &      9 &     10 &     10 &     10 &      10 &      10 \\
FCNN SQGL NoSyn PosNeg &      0 &      1 &      0 &      0 &       1 &       0 \\
FCNN SQGL Syn PosNeg &      7 &     10 &     10 &     10 &      10 &      10 \\
\bottomrule
\end{tabular}

\begin{tabular}{lrrrrrr}
\toprule
&&USPS \\
\midrule
                     & 1-tree & 2-tree & 4-tree & 8-tree & 16-tree & 32-tree \\
\midrule
k-tree SQGL NoSyn PosNeg &      8 &      6 &      7 &     10 &      10 &      10 \\
k-tree SQGL Syn PosNeg &     10 &     10 &     10 &     10 &      10 &      10 \\
FCNN SQGL NoSyn PosNeg &     10 &     10 &     10 &     10 &      10 &      10 \\
FCNN SQGL Syn PosNeg &     10 &     10 &     10 &     10 &      10 &      10 \\
\bottomrule
\end{tabular}

\begin{tabular}{lrrrrrr}
\toprule
&&CIFAR10 \\
\midrule
                     & 1-tree & 2-tree & 4-tree & 8-tree & 16-tree & 32-tree \\
\midrule
k-tree SQGL NoSyn PosNeg &     10 &      9 &      8 &      8 &       9 &       8 \\
k-tree SQGL Syn PosNeg &      6 &      9 &     10 &      9 &       9 &      10 \\
FCNN SQGL NoSyn PosNeg &      8 &      8 &     10 &     10 &      10 &      10 \\
FCNN SQGL Syn PosNeg &      7 &      7 &      3 &      5 &       7 &       3 \\
\bottomrule
\end{tabular}
\end{center}
\caption{Trial Numbers for Figure \ref{fig:synapse}, Comparing $k$-tree and FCNN models with or without Synapse Nonlinearity.}
\label{tab-synapse}
\end{table}

\begin{table}[ht]
\tiny
\begin{center}
\begin{tabular}{lrrrrrr}
\toprule
&&MNIST \\
\midrule
                      & 1-tree & 2-tree & 4-tree & 8-tree & 16-tree & 32-tree \\
\midrule
k-tree NaCaK NoSyn PosNeg &     10 &     10 &     10 &     10 &      10 &      10 \\
k-tree NaCaK NoSyn Pos &      5 &      9 &     10 &     10 &      10 &      10 \\
k-tree NaCaK Syn PosNeg &      9 &     10 &     10 &     10 &      10 &      10 \\
k-tree NaCaK Syn Pos &     10 &     10 &     10 &     10 &      10 &      10 \\
FCNN NaCaK NoSyn PosNeg &     10 &     10 &     10 &     10 &      10 &      10 \\
FCNN NaCaK Syn PosNeg &     10 &     10 &     10 &     10 &      10 &      10 \\
\bottomrule
\end{tabular}

\begin{tabular}{lrrrrrr}
\toprule
&&FMNIST \\
\midrule
                      & 1-tree & 2-tree & 4-tree & 8-tree & 16-tree & 32-tree \\
\midrule
k-tree NaCaK NoSyn PosNeg &      9 &     10 &     10 &     10 &      10 &      10 \\
k-tree NaCaK NoSyn Pos &      4 &     10 &     10 &     10 &      10 &      10 \\
k-tree NaCaK Syn PosNeg &     10 &     10 &     10 &     10 &      10 &      10 \\
k-tree NaCaK Syn Pos &     10 &     10 &     10 &     10 &      10 &      10 \\
FCNN NaCaK NoSyn PosNeg &     10 &     10 &     10 &     10 &      10 &      10 \\
FCNN NaCaK Syn PosNeg &     10 &     10 &     10 &     10 &      10 &      10 \\
\bottomrule
\end{tabular}

\begin{tabular}{lrrrrrr}
\toprule
&&KMNIST \\
\midrule
                      & 1-tree & 2-tree & 4-tree & 8-tree & 16-tree & 32-tree \\
\midrule
k-tree NaCaK NoSyn PosNeg &      9 &     10 &     10 &     10 &      10 &      10 \\
k-tree NaCaK NoSyn Pos &      6 &      8 &     10 &     10 &      10 &      10 \\
k-tree NaCaK Syn PosNeg &     10 &     10 &     10 &     10 &      10 &      10 \\
k-tree NaCaK Syn Pos &     10 &     10 &     10 &     10 &      10 &      10 \\
FCNN NaCaK NoSyn PosNeg &     10 &     10 &     10 &     10 &      10 &      10 \\
FCNN NaCaK Syn PosNeg &     10 &     10 &     10 &     10 &      10 &      10 \\
\bottomrule
\end{tabular}

\begin{tabular}{lrrrrrr}
\toprule
&&EMNIST \\
\midrule
                      & 1-tree & 2-tree & 4-tree & 8-tree & 16-tree & 32-tree \\
\midrule
k-tree NaCaK NoSyn PosNeg &     10 &     10 &     10 &     10 &      10 &      10 \\
k-tree NaCaK NoSyn Pos &     10 &     10 &     10 &     10 &      10 &      10 \\
k-tree NaCaK Syn PosNeg &     10 &     10 &     10 &     10 &      10 &      10 \\
k-tree NaCaK Syn Pos &     10 &     10 &     10 &     10 &      10 &      10 \\
FCNN NaCaK NoSyn PosNeg &     10 &     10 &     10 &     10 &      10 &      10 \\
FCNN NaCaK Syn PosNeg &     10 &     10 &     10 &     10 &      10 &      10 \\
\bottomrule
\end{tabular}

\begin{tabular}{lrrrrrr}
\toprule
&&SVHN \\
\midrule
                      & 1-tree & 2-tree & 4-tree & 8-tree & 16-tree & 32-tree \\
\midrule
k-tree NaCaK NoSyn PosNeg &      2 &      3 &      4 &      4 &       3 &       2 \\
k-tree NaCaK NoSyn Pos &      0 &      1 &      0 &      0 &       0 &       0 \\
k-tree NaCaK Syn PosNeg &      9 &     10 &     10 &     10 &      10 &      10 \\
k-tree NaCaK Syn Pos &     10 &      9 &     10 &     10 &      10 &      10 \\
FCNN NaCaK NoSyn PosNeg &      0 &      1 &      0 &      0 &       1 &       0 \\
FCNN NaCaK Syn PosNeg &      7 &     10 &     10 &     10 &      10 &      10 \\
\bottomrule
\end{tabular}

\begin{tabular}{lrrrrrr}
\toprule
&&USPS \\
\midrule
                      & 1-tree & 2-tree & 4-tree & 8-tree & 16-tree & 32-tree \\
\midrule
k-tree NaCaK NoSyn PosNeg &      8 &      6 &      7 &     10 &      10 &      10 \\
k-tree NaCaK NoSyn Pos &      1 &     10 &     10 &     10 &      10 &      10 \\
k-tree NaCaK Syn PosNeg &     10 &     10 &     10 &     10 &      10 &      10 \\
k-tree NaCaK Syn Pos &     10 &     10 &     10 &     10 &      10 &      10 \\
FCNN NaCaK NoSyn PosNeg &     10 &     10 &     10 &     10 &      10 &      10 \\
FCNN NaCaK Syn PosNeg &     10 &     10 &     10 &     10 &      10 &      10 \\
\bottomrule
\end{tabular}

\begin{tabular}{lrrrrrr}
\toprule
&&CIFAR10 \\
\midrule
                      & 1-tree & 2-tree & 4-tree & 8-tree & 16-tree & 32-tree \\
\midrule
k-tree NaCaK NoSyn PosNeg &     10 &      9 &      8 &      8 &       9 &       8 \\
k-tree NaCaK NoSyn Pos &      0 &      1 &      4 &      2 &       3 &       3 \\
k-tree NaCaK Syn PosNeg &      6 &      9 &     10 &      9 &       9 &      10 \\
k-tree NaCaK Syn Pos &      9 &     10 &     10 &     10 &       9 &      10 \\
FCNN NaCaK NoSyn PosNeg &      8 &      8 &     10 &     10 &      10 &      10 \\
FCNN NaCaK Syn PosNeg &      7 &      7 &      3 &      5 &       7 &       3 \\
\bottomrule
\end{tabular}
\end{center}
\caption{Trial Numbers for Figure \ref{fig:constantsqgl}, Comparing $k$-tree and FCNN models with or without Non-negative Weight Constraint.}
\label{tab-constantsqgl}
\end{table}

\begin{table}[ht]
\tiny
\begin{center}
\begin{tabular}{lrrrrrr}
\toprule
&&&MNIST \\
\midrule
      & 1-tree & 2-tree & 4-tree & 8-tree & 16-tree & 32-tree \\
\midrule
ReLU &     10 &     10 &     10 &     10 &      10 &      10 \\
LReLU &     10 &     10 &     10 &     10 &      10 &      10 \\
Sigmoid &     10 &     10 &     10 &     10 &      10 &      10 \\
Swish &     10 &     10 &     10 &     10 &      10 &      10 \\
NaCaK &     10 &     10 &     10 &     10 &      10 &      10 \\
\bottomrule
\end{tabular}

\begin{tabular}{lrrrrrr}
\toprule
&&&FMNIST \\
\midrule
      & 1-tree & 2-tree & 4-tree & 8-tree & 16-tree & 32-tree \\
\midrule
ReLU &      7 &     10 &     10 &     10 &      10 &      10 \\
LReLU &     10 &     10 &     10 &     10 &      10 &      10 \\
Sigmoid &     10 &     10 &     10 &     10 &      10 &      10 \\
Swish &     10 &     10 &     10 &     10 &      10 &      10 \\
NaCaK &     10 &     10 &     10 &     10 &      10 &      10 \\
\bottomrule
\end{tabular}

\begin{tabular}{lrrrrrr}
\toprule
&&&KMNIST \\
\midrule
      & 1-tree & 2-tree & 4-tree & 8-tree & 16-tree & 32-tree \\
\midrule
ReLU &     10 &     10 &     10 &     10 &      10 &      10 \\
LReLU &     10 &     10 &     10 &     10 &      10 &      10 \\
Sigmoid &     10 &     10 &     10 &     10 &      10 &      10 \\
Swish &     10 &     10 &     10 &     10 &      10 &      10 \\
NaCaK &     10 &     10 &     10 &     10 &      10 &      10 \\
\bottomrule
\end{tabular}

\begin{tabular}{lrrrrrr}
\toprule
&&&EMNIST \\
\midrule
      & 1-tree & 2-tree & 4-tree & 8-tree & 16-tree & 32-tree \\
\midrule
ReLU &     10 &     10 &     10 &     10 &      10 &      10 \\
LReLU &     10 &     10 &     10 &     10 &      10 &      10 \\
Sigmoid &     10 &     10 &     10 &     10 &      10 &      10 \\
Swish &     10 &     10 &     10 &     10 &      10 &      10 \\
NaCaK &     10 &     10 &     10 &     10 &      10 &      10 \\
\bottomrule
\end{tabular}

\begin{tabular}{lrrrrrr}
\toprule
&&&SVHN \\
\midrule
      & 1-tree & 2-tree & 4-tree & 8-tree & 16-tree & 32-tree \\
\midrule
ReLU &      1 &      0 &      1 &      4 &       2 &       5 \\
LReLU &      4 &      6 &      8 &      6 &       7 &       9 \\
Sigmoid &      1 &      0 &      1 &      0 &       2 &       1 \\
Swish &      2 &      2 &      0 &      1 &       0 &       2 \\
NaCaK &      0 &      1 &      0 &      0 &       1 &       0 \\
\bottomrule
\end{tabular}

\begin{tabular}{lrrrrrr}
\toprule
&&&USPS \\
\midrule
      & 1-tree & 2-tree & 4-tree & 8-tree & 16-tree & 32-tree \\
\midrule
ReLU &      8 &     10 &     10 &     10 &      10 &      10 \\
LReLU &     10 &     10 &     10 &     10 &      10 &      10 \\
Sigmoid &     10 &     10 &     10 &     10 &      10 &      10 \\
Swish &     10 &     10 &     10 &     10 &      10 &      10 \\
NaCaK &     10 &     10 &     10 &     10 &      10 &      10 \\
\bottomrule
\end{tabular}

\begin{tabular}{lrrrrrr}
\toprule
&&&CIFAR10 \\
\midrule
      & 1-tree & 2-tree & 4-tree & 8-tree & 16-tree & 32-tree \\
\midrule
ReLU &      0 &      1 &      1 &      2 &       2 &       3 \\
LReLU &      9 &      9 &      9 &     10 &       7 &       8 \\
Sigmoid &      5 &      6 &      4 &      6 &       6 &       3 \\
Swish &      5 &      4 &      5 &      4 &       6 &       3 \\
NaCaK &      8 &      8 &     10 &     10 &      10 &      10 \\
\bottomrule
\end{tabular}
\end{center}
\caption{Trial Numbers for Figure \ref{fig:fcnn_nonlins}, Comparing Nonlinear Activation Functions for the Dendrite Nodes in the FCNN For Each Dataset.}
\label{tab-fcnn_nonlins}
\end{table}

\end{document}